\documentclass[a4paper,fleqn,usenatbib]{mnras}

\usepackage{newtxtext,newtxmath}

\usepackage[T1]{fontenc}
\usepackage{ae,aecompl}

\usepackage{graphicx}	
\usepackage{amsmath}	
\usepackage{amssymb}	
\usepackage{ulem, epstopdf, mathtools, amsfonts, multirow, makecell}

\newcommand{\rmd}{\,\mathrm{d}}
\newcommand{\rmi}{\,\mathrm{i}}
\newcommand{\del}{\partial}

\title[Effects of strong magnetic field on gravity waves]{Effects of a strong magnetic field on internal gravity waves: trapping, phase mixing, reflection and dynamical chaos}

\author[Loi \& Papaloizou]{
Shyeh Tjing Loi\thanks{E-mail: stl36@cam.ac.uk} and
John C. B. Papaloizou\thanks{E-mail: jcbp2@damtp.cam.ac.uk}
\\
Department of Applied Mathematics and Theoretical Physics, University of Cambridge, Centre for Mathematical Sciences, Wilberforce Road, Cambridge CB3 0WA, UK
}

\date{Accepted: ??? Received: ???; in original form: ???}

\pubyear{2018}

\begin{document}
\label{firstpage}
\pagerange{\pageref{firstpage}--\pageref{lastpage}}
\maketitle

\begin{abstract}
  The spectrum of oscillation modes of a star provides information not only about its material properties (e.g.~mean density), but also its symmetries. Spherical symmetry can be broken by rotation and/or magnetic fields. It has been postulated that strong magnetic fields in the cores of some red giants are responsible for their anomalously weak dipole mode amplitudes (the ``dipole dichotomy'' problem), but a detailed understanding of how gravity waves interact with strong fields is thus far lacking. In this work, we attack the problem through a variety of analytical and numerical techniques, applied to a localised region centred on a null line of a confined axisymmetric magnetic field which is approximated as being cylindrically symmetric. We uncover a rich variety of phenomena that manifest when the field strength exceeds a critical value, beyond which the symmetry is drastically broken by the Lorentz force. When this threshold is reached, the spatial structure of the g-modes becomes heavily altered. The dynamics of wave packet propagation transitions from regular to chaotic, which is expected to fundamentally change the organisation of the mode spectrum. In addition, depending on their frequency and the orientation of field lines with respect to the stratification, waves impinging on different parts of the magnetised region are found to undergo either reflection or trapping. Trapping regions provide an avenue for energy loss through Alfv\'{e}n wave phase mixing. Our results may find application in various astrophysical contexts, including the dipole dichotomy problem, the solar interior, and compact star oscillations.
\end{abstract}




\section{Introduction}
Stars are self-gravitating fluid bodies which are spherically symmetric only to first approximation \citep[e.g.][]{Perdang1986}. Many processes act to break this spherical symmetry: there are those occuring on local scales, such as convection, granulation and turbulence, and those on global scales, such as tides, rotation and magnetism. Oscillations in stars, formed by constructive interference of propagating waves, have a spatial and temporal organisation that reflects not just local fluid mechanical properties, but also global symmetries. By Noether's theorem, symmetries are fundamentally associated with conserved quantities. Such concepts are pertinent to the Hamiltonian treatment of classical mechanics, a framework applicable to fluid mechanics in the Wentzel-Kramers-Brillouin (WKB) or short-wavelength limit. In this scenario, Hamiltonian trajectories correspond to the propagation paths of wave packets. A Hamiltonian system for which there exist as many conserved quantities (actions) as spatial degrees of freedom, for example wave packets in a non-evolving, perfectly spherical star, is said to be integrable. Such systems have the property that they cannot be chaotic, and in the absence of singularities straightforward methods exist for obtaining the spectrum of normal modes in the short-wavelength limit by applying a set of quantisation conditions demanding constructive interference of propagating waves \citep{Keller1960, Gough1986}. If the symmetry is only weakly broken, the dynamical properties of such a system can be regarded as small perturbations about the integrable case, and standard results/asymptotic formulae \citep[e.g.][]{Tassoul1980} remain largely valid.

For systems in which the symmetry is significantly broken, such as in tidally/centrifugally distorted stars and/or those with strong magnetic fields, the wave packet dynamics will undergo a transition to chaos. Such concepts, first developed and applied in the field of quantum mechanics \citep{Hentschel2002, Gutzwiller1990, Cao2015}, have in recent years been used to investigate the effects of rapid stellar rotation on acoustic \citep{Lignieres2008, Lignieres2009, Lignieres2010} and gravity modes \citep{Prat2016, Prat2017, Prat2018}. These studies show that associated with the breaking of spherical symmetry is the appearance of chaotic modes, which coexist alongside regular modes and occupy a larger and larger volume of phase space as the rotation rate increases. Such modes have fundamentally different properties from regular modes in that their frequencies do not obey a predictable pattern (the conserved actions in terms of which the eigenspectrum might be defined do not exist), and can only be characterised in a statistical sense. In general, the normal mode spectrum of a non-integrable system is a superposition of regular and irregular spectra \citep{Percival1973, Pomphrey1974}. The evolution of an integrable system towards chaos under a symmetry-breaking perturbation occurs in a smooth manner, as described by the Kolmogorov-Arnold-Moser (KAM) theorem \citep{Kolmogorov1954, Arnold1963, Moser1962}. Although the existence of any such perturbation (however small) leads to chaos, the associated volume of phase space vanishes in the limit of zero perturbation, making this undetectable for near-integrable systems. The observation that many stars exhibit predominantly regular spectra implies that they are spherically symmetric to good approximation, and thus fall under this category.

While the study of dynamical chaos has been undertaken in the context of wave packets propagating in rapidly rotating stars, it has not yet been attempted for the problem of strong magnetic fields. Generally speaking, the magnetic field problem is less straightforward to approach, owing to the fact that physically realistic field configurations are difficult to analytically obtain and parametrise \citep[e.g.][]{Lyutikov2010}. This is compounded by the lack of reliable methods for probing the magnetic field beneath the stellar surface. Note that unlike for acoustic waves, it is not necessary for the magnetic and gas pressures to be comparable for efficient interactions to occur (indeed, the plasma $\beta$ is much greater than unity in typical stellar interiors). At any given non-zero field strength, interactions with gravity waves are always possible in some part of phase space, this shifting to smaller spatial scales when fields are weaker. The strong-field and rapid-rotation problems share a commonality in that the effects on the normal mode spectrum go beyond what can be described by first-order non-degenerate perturbation theory (see, e.g., \citet{Reese2010} for a discussion of the rotation problem). The attempted application of first-order perturbation theory by \citet{Cantiello2016} to predicting the magnetic splitting associated with strong core fields in red giants found that this is at least comparable to the frequency spacing of modes in the non-magnetic case, which supports this notion. This, and the numerical work of \citet{Lecoanet2017}, which shows significant acquisition of Alfv\'{e}nic character by gravity waves in a strong field, point to substantial qualitative differences between modes present in an unmagnetised star and one that is strongly magnetised in at least part of its volume. However, consequences for global (mode) dynamics in stars with strong fields have yet to be established at any convincing level. To make further progress, it is necessary to break away from the techniques and paradigms of standard stellar oscillation theory, which have been formulated around the assumption of spherical symmetry, and adopt a more general approach.

The goal of the current work is to investigate further the physical properties and interactions of gravity waves/modes and magnetic fields of dynamically significant strengths. It is expected that gravity waves will begin to interact with magnetic fields in a dynamically important way when Alfv\'{e}n frequencies become comparable to those of gravity waves, for disturbances of a given spatial scale. We approach this using a combination of two complementary methods, namely (i) nonlinear Boussinesq simulations using the spectral code \textsc{snoopy} \citep{Lesur2007, Lesur2015}, and (ii) Hamiltonian ray tracing. Our model, simulation setup and ray-tracing method are described in \S\ref{sec:models}. In \S\ref{sec:results} we describe the variety of processes associated with the interaction, including global modifications to the allowed form of g-modes arising from reflection at the top of the magnetic region and trapping within it. In particular, we relate results obtained from the simulations to the properties of propagating wave packets. We discuss the broader implications and state the limitations of our findings in \S\ref{sec:discuss}. Finally, we conclude in \S\ref{sec:summary}.

\begin{figure}
  \centering
  \includegraphics[width=\columnwidth]{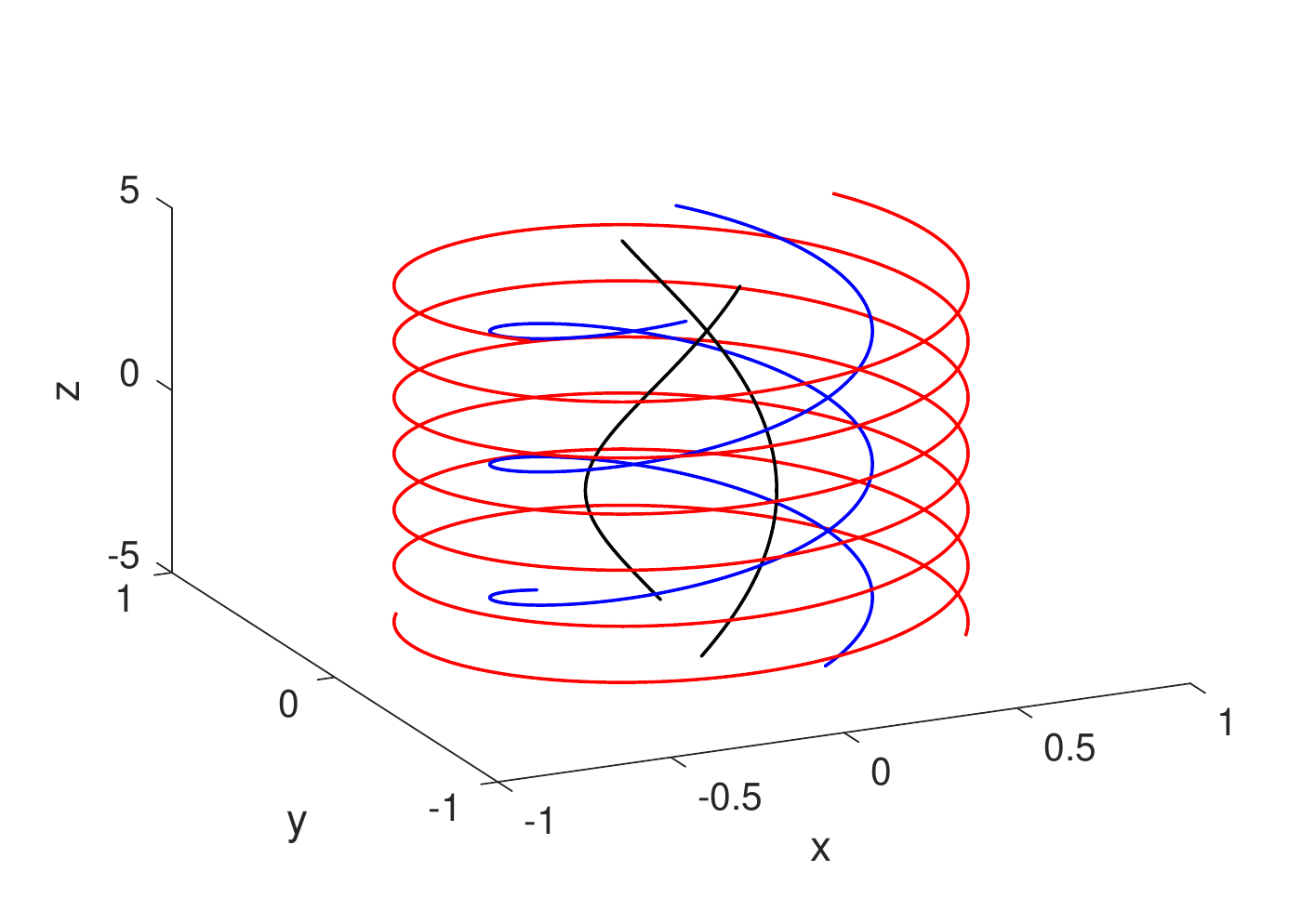}
  \includegraphics[width=0.85\columnwidth]{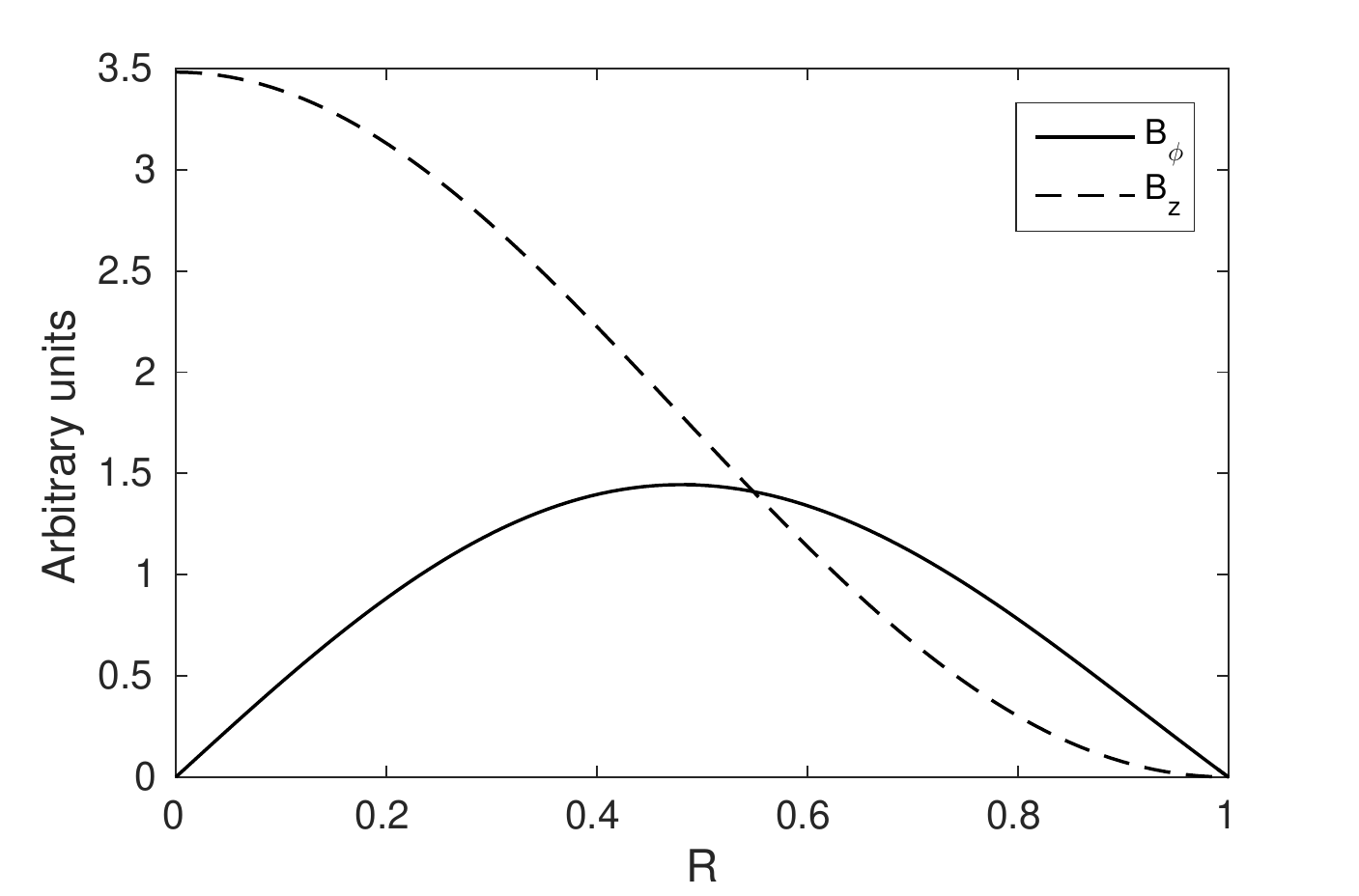}
  \caption{Top: Representative field lines at radial distances of 0.25 (black), 0.5 (blue) and 0.75 (red) units from the $z$-axis. The radius of the cylinder is $a = 1$. Bottom: The $\phi$ and $z$-components of the field, as a function of radius given by Equation (\ref{eq:B_soln}).}
  \label{fig:cylPrendergast}
\end{figure}

\section{Models \& methods}\label{sec:models}
Since the goal of this work is to elucidate the nature of the physical processes involved, we adopt a fairly uncomplicated geometry that is straightforward to implement and analyse. The setup consists of a Cartesian box stratified along a single direction (the $x$-direction), in which a helical magnetic field is embedded occupying a cylindrical volume whose axis is orthogonal to the stratification. We take the axis of the cylinder to be the $z$-direction. The field configuration (details below) is such that all three field components, which are each in general non-zero within the cylinder, vanish at the boundary of the cylinder, whose diameter is set to be half the width of the box. This is intended to be the cylindrical analogue of the Prendergast solution used in our previous work \citep[][which we henceforth abbreviate to LP17]{Loi2017}. The whole setup is translationally symmetric in the $z$-direction, which is to be associated with the azimuthal direction in a spherical star (the symmetry direction, since the Prendergast field is axisymmetric). As for the spherical case, in the absence of a magnetic field the system possesses additional spatial symmetries (here, translational symmetry in the $x$ and $y$ directions), but these are broken when the field is non-zero.

\subsection{Cylindrical Prendergast solution}\label{sec:cylPrendergast}
To model a magnetic field embedded within a star, one desires an equilibrium solution that is non-singular, vanishes smoothly at a finite distance from the origin, and that has a mixed poloidal-toroidal topology. The last property is required for the sake of dynamical stability, as established in previous studies \citep{Tayler1973, Flowers1977, Braithwaite2006}. For spherical, incompressible stars, an analytic solution with these properties was found by \citet{Prendergast1956}, derived by considering the Grad-Shafranov equation and assuming near-barotropy. This was later extended by \citet{Duez2010, Duez2010a} to incorporate compressibility. We refer to the generalised (compressible) result as the \textit{Prendergast solution}. To adapt this for the Cartesian geometry used in this work, a topologically equivalent, cylindrical version of the Prendergast solution will be derived using an approach analogous to the spherical case (for details of the latter, refer to LP17, section 2.2). 

In the following, we adopt cylindrical polar coordinates $(R, \phi, z)$. The solution we seek is an axisymmetric one with translational symmetry along $z$, i.e.~a circular helix whose pitch angle varies only with $R$. Such a magnetic field configuration can be expressed in the general form
\begin{align}
  \mathbf{B} = \nabla \times (\psi \mathbf{\hat{z}}) + B_z \mathbf{\hat{z}} \:, \label{eq:fluxfn}
\end{align}
where $\psi$ is a scalar encoding information about the $x$- and $y$-components of $\mathbf{B}$ (analogues of the poloidal components) and $B_z$ is the analogue of the toroidal component. See that under $\del/\del\phi \equiv 0$, Equation (\ref{eq:fluxfn}) corresponds to a field with no $R$-component. As we have $\del/\del z \equiv 0$, it follows that $\mathbf{B} \cdot \nabla \psi = 0$, meaning that $\psi$ is invariant on field lines and can be interpreted as a flux function.

Considering that $[(\nabla \times \mathbf{B}) \times \mathbf{B}] \cdot \hat{\mathbf{z}} = 0$, one can show with aid of appropriate vector identities that $B_z = B_z(\psi)$. Following \citet{Prendergast1956}, it is convenient to choose $B_z(\psi) = \lambda \psi$, where $\lambda$ is a global constant. In addition, if one substitutes Equation (\ref{eq:fluxfn}) into the force-balance condition
\begin{align}
  \nabla p + \rho \nabla \Phi = (\nabla \times \mathbf{B}) \times \mathbf{B} \:, \label{eq:MHS}
\end{align}
with $p$ being pressure, $\rho$ mass density and $\Phi$ the gravitational potential, and assumes a barotropic equation of state, this leads to the linear second-order ordinary differential equation (ODE)
\begin{align}
  \nabla^2 \psi + \lambda^2 \psi = \kappa \rho \:. \label{eq:GSE}
\end{align}
Under the above assumptions, one can show $\kappa$ to be a constant on flux surfaces. After \citet{Prendergast1956}, we in addition choose to this be a global constant. In the spirit of the Boussinesq approximation (see later) $\rho$ will also be taken as constant, equal to unity in our system of units. The scaling of the magnetic field has been chosen such that the quantity $\mu_0$ that usually appears in the Lorentz force is replaced by unity. Equation (\ref{eq:GSE}) is the cylindrical analogue of the Grad-Shafranov equation applying to spherical stars, and in this case can be recognised as the inhomogeneous Bessel differential equation.

Let the outer radius of the axisymmetric magnetic cylinder be $a$. The boundary conditions we wish to impose are
\begin{align}
  B_\phi(0) &= 0 \:, \quad B_\phi(a) = 0 \:, \quad B_z(a) = 0 \:, \label{eq:BCs} \\
  \shortintertext{which, given that $\mathbf{B} = (0, -\psi', \lambda \psi)$, translate to}
  \psi'(0) &= 0 \:, \quad \psi'(a) = 0 \:, \quad \psi(a) = 0 \:. \label{eq:psi_BCs}
\end{align}
A prime here denotes a derivative with respect to $R$. It is straightforward to check that the solution
\begin{align}
  \psi(R) = \frac{\kappa \rho}{\lambda^2} \left( 1 - \frac{J_0(\lambda R)}{J_0(\lambda a)} \right) \:, \label{eq:psi_soln}
\end{align}
with $\lambda$ such that $J_1(\lambda a) = 0$, satisfies Equations (\ref{eq:GSE}) and (\ref{eq:psi_BCs}), where $J_0$ and $J_1$ are Bessel functions of the first kind. The magnetic field solution for $R < a$ is then
\begin{align}
  \mathbf{B} = \frac{\kappa \rho}{\lambda} \left(0, -\frac{J_1(\lambda R)}{J_0(\lambda a)}, 1 - \frac{J_0(\lambda R)}{J_0(\lambda a)} \right) \:, \label{eq:B_soln}
\end{align}
matched to a zero solution for $R \geq a$. This is plotted in Figure \ref{fig:cylPrendergast}.

A notable qualitative difference between the cylindrical and spherical Prendergast solutions is that while for an incompressible fluid there exists in the spherical case a coupling between axisymmetric spheroidal and torsional motions through the Lorentz force, this coupling vanishes when the radial distance from the axis of symmetry is taken to infinity (cylindrical limit). In our setup, the $x,y$-plane is analogous to the meridional plane of a sphere and the $z$-axis is analogous to the azimuthal direction. For the spherical Prendergast solution, the Lorentz force perturbation associated with purely spheroidal motions has, to first order, a non-zero component in the torsional direction. However, for the cylindrical Prendergast solution, the $z$-component of the Lorentz force perturbation associated with purely $x,y$-motions is zero to first order.

To see this, let us begin by considering the perturbation to the Lorentz force by axisymmetric disturbances in the spherical case, denoting the azimuthal (torsional) direction by $\varphi$ and distance from the axis of symmetry by $r$. For an axisymmetric background field, the magnetic pressure term makes no contribution to the $\varphi$-component of the Lorentz force, and so we are left to examine the magnetic tension term,
\begin{align}
  \left[ (\mathbf{B} \cdot \nabla) \mathbf{B} \right]'_\varphi = (\mathbf{B}_0 \cdot \nabla) B'_\varphi + (\mathbf{B}' \cdot \nabla) B_{0\varphi} + \frac{B'_\varphi B_{0r}}{r} + \frac{B_{0\varphi} B'_r}{r} \:. \label{eq:spher_tension}
\end{align}
Subscript 0's denote background quantities, and primes denote perturbations about the background. The cylindrical Prendergast solution corresponds to the limit where $r \to \infty$, with $\varphi$ replaced by the Cartesian coordinate $z$. The last two terms on the right-hand side of Equation (\ref{eq:spher_tension}), which are associated with spatial curvature of the coordinates, thus vanish. If in addition we impose incompressibility ($\nabla \cdot \boldsymbol{\xi} = 0$, where $\boldsymbol{\xi}$ is the fluid displacement), and eliminate all other perturbed quantities in favour of $\boldsymbol{\xi}$ using the linearised induction equation $\mathbf{B}' = \nabla \times (\boldsymbol{\xi} \times \mathbf{B}_0)$, we find that in the limit $r \to \infty$
\begin{align}
  \left[ (\mathbf{B} \cdot \nabla) \mathbf{B} \right]'_\varphi \to \left[ (\mathbf{B} \cdot \nabla) \mathbf{B} \right]'_z = -\boldsymbol{\xi} \cdot \nabla (\mathbf{B}_0 \cdot \nabla B_{0z}) \:. \label{eq:cyl_tension}
\end{align}
This equals zero, since $\mathbf{B}_0 \cdot \nabla B_{0z}$ corresponds to the $z$-component of the magnetic tension associated with the background (equilibrium) field. No other forces act in this direction, and so it must be zero.

It would be desirable to be able to investigate the mutual excitation of (the analogues to) spheroidal and torsional modes discussed in LP17, using our nonlinear Boussinesq simulations. Note that this coupling in general vanishes in the limit of zero wavelength, but is present for finite wavelengths when the appropriate curvature in the system exists. Given that the cylindrical Prendergast field configuration does not replicate this property of the spherical version owing to the lack of curvature in the symmetry direction, this coupling between $x,y$ motions and the $z$ direction must be artificially induced. We achieve this through suitable modifications to the numerical code, described further below. 

\begin{figure*}
  \centering
  \includegraphics[clip=true, trim=2cm 0cm 2cm 0cm, width=\textwidth]{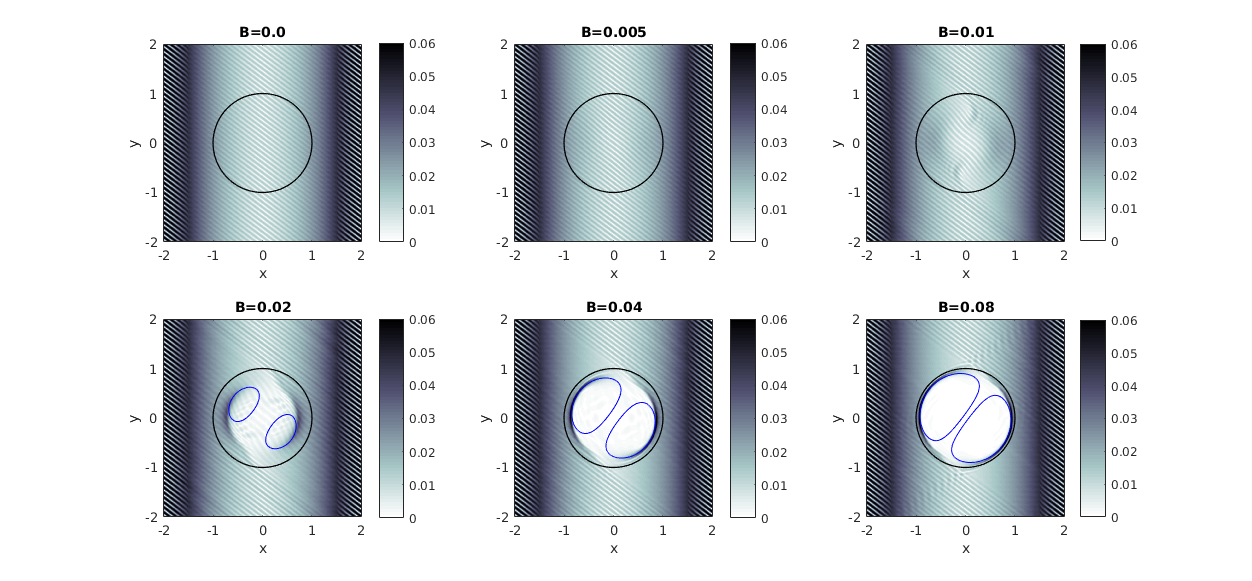}
  \caption{Spatial distribution of the temporal Fourier amplitude at the forcing frequency ($\omega_f = 0.8$) for the $v_y$ field, at the six values of the field strength simulated with \textsc{snoopy}. The Fourier transform is taken over the interval $200 < t < 1000$. The forced mode has $[i,j,k]=[15,20,0]$ (steep), and the forcing region lies in the strip outside $|x| > 1.5$. The black circle of radius 1 marks the boundary of the field region, and overlaid in blue are the critical surfaces (i.e., where $\omega_f = \omega_A \equiv \mathbf{k} \cdot \mathbf{v}_A$, $\mathbf{k}$ in this case being the forced wavenumber). The critical field strength here is $B \sim 0.02$.}
  \label{fig:steep_vy_tFT}
\end{figure*}

\begin{figure*}
  \centering
  \includegraphics[clip=true, trim=2cm 0cm 2cm 0cm, width=\textwidth]{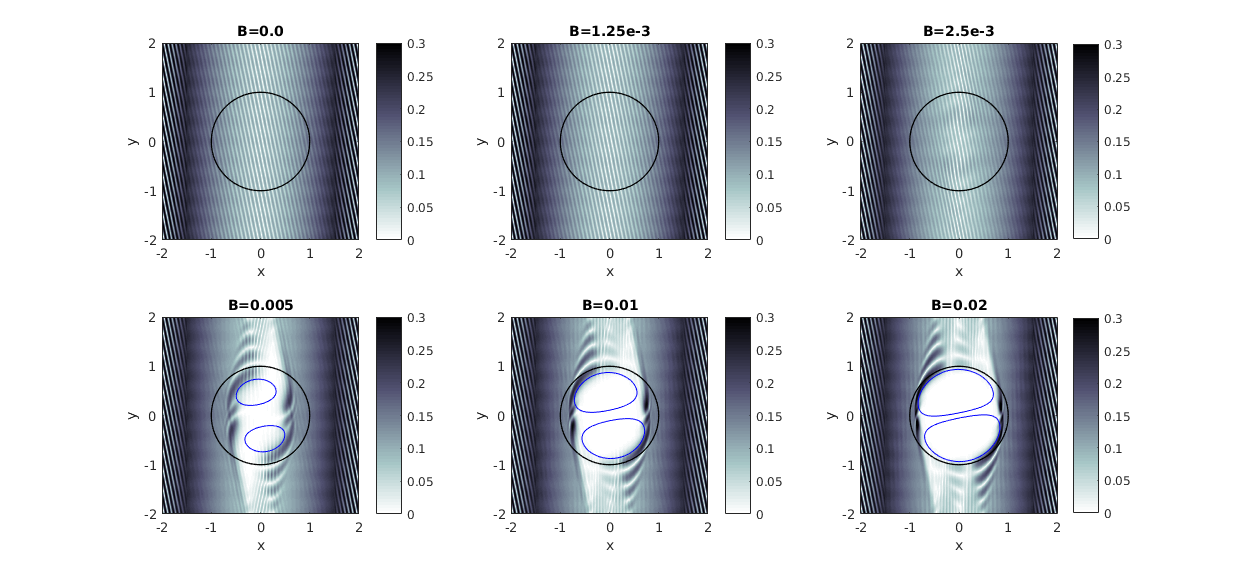}
  \caption{As for Fig.~\ref{fig:steep_vy_tFT}, but for the \textsc{snoopy} simulation with $[i,j,k]=[25,5,0]$ (shallow). The forcing frequency in this case is $\omega_f = 0.196$, and the critical field strength is $B \sim 0.005$.}
  \label{fig:shallow_vy_tFT}
\end{figure*}

\subsection{MHD Boussinesq simulations}\label{sec:snoopy}
The incompressible 3D spectral code \textsc{snoopy} \citep{Lesur2007, Lesur2015} was used to solve the dissipative MHD Boussinesq equations, given in non-dimensional form by
\begin{align}
  \frac{\del \mathbf{u}}{\del t} + \mathbf{u} \cdot \nabla \mathbf{u} &= -\nabla \Pi + N^2 \theta \,\hat{\mathbf{x}} + \mathbf{B} \cdot \nabla \mathbf{B} + \frac{\nabla^2 \mathbf{u}}{\mathrm{Re}} \label{eq:mom} \\
  \frac{\del \mathbf{B}}{\del t} &= \nabla \times (\mathbf{u} \times \mathbf{B}) + \frac{\nabla^2 \mathbf{B}}{\mathrm{Rm}} \label{eq:ind} \\
  \frac{\del \theta}{\del t} + \mathbf{u} \cdot \nabla \theta &= -u_x + \frac{\nabla^2 \theta}{\mathrm{Rt}} \label{eq:therm} \\
  \nabla \cdot \mathbf{u} &= 0 \:, \quad \nabla \cdot \mathbf{B} = 0 \:. \label{eq:div0}
\end{align}
Here $\mathbf{u}$ is the fluid velocity, $\theta$ is the potential temperature, $N^2$ is the squared buoyancy frequency, $\Pi$ is the combined gas and magnetic pressure, Re is the Reynolds number, Rm is the magnetic Reynolds number, and Rt is a non-dimensional number controlling the thermal diffusivity (the ``thermal Reynolds number''). The fluid is stratified in the $x$-direction, and $N^2 = 1$ (in our system of units) is taken to be a global constant. While a constant buoyancy frequency may at first glance seem an adverse simplification, note that it is the relative value of the buoyancy and Alfv\'{e}n frequency which is important in controlling the strength of coupling between gravity and Alfv\'{e}n waves. The latter we do vary substantially through our field configuration: the field amplitude is zero within most of the box, and attains a value inside the cylinder that is large enough to affect inwardly propagating g-modes.

Given that \textsc{snoopy} is a spectral code, all boundary conditions are necessarily periodic. The simulation domain is a Cartesian box with dimensions $[L_x,L_y,L_z]=[4,4,1]$ in our system of units, with a uniform grid in each direction having a resolution of $1024 \times 1024 \times 8$. We remark that in this paper we study motions that are independent of $z$, hence the choice of a small number of grid points in that direction. The magnetic field is confined to a cylinder with radius $a = 1$ centred within the box and aligned along the $z$-axis. The width of the box effectively sets the length scale of the problem, while the buoyancy frequency sets the time scale. Note that this local box model follows from the general axisymmetric case in the high-wavenumber limit, but applied to a special field configuration that is a slender torus encircling the equator. While a slender-torus configuration may not be fully realistic, we shall argue later in \S\ref{sec:discuss} that the main results of this work are still generalisable to short-wavelength waves in the full spherical case.

In the absence of a magnetic field, the normal modes are plane waves having an integer number of undulations within the box in each of the $x$, $y$ and $z$ directions. It is convenient to label these modes by a triplet of integers $[i,j,k]$ representing the number of wavelengths spanned in each direction: for example, $[i,j,k] = [4,5,2]$ denotes a plane wave with four undulations across the box in the $x$-direction, five in the $y$-direction and two in the $z$-direction. The associated wavevector is given by
\begin{align}
  \mathbf{k} = \left( k_x, k_y, k_z \right) = 2\pi \left( \frac{i}{L_x}, \frac{j}{L_y}, \frac{k}{L_z} \right) \:. \label{eq:kvec}
\end{align}
These modes are gravity modes, and so the frequency $\omega$ is given by the dispersion relation for gravity waves
\begin{align}
  \omega^2 &= \kappa_\perp^2 N^2 \:, \label{eq:gravity_DR} \\
  \shortintertext{where}
  \kappa_\perp^2 &\equiv \frac{k_y^2 + k_z^2}{k_x^2 + k_y^2 + k_z^2} \label{eq:kappa_perp}
\end{align}
is the square of the component of the unit vector perpendicular to the stratification.

In solar-like oscillators, the source of energy for gravity wave excitation lies in the convective envelope, outside the stably-stratified, possibly magnetised region \citep{Andersen1996, Rogers2005, Belkacem2009, Alvan2014}. To model this in our \textsc{snoopy} simulations, we applied a continuous forcing to the region $|x| > 1.5$, which lies exterior to the field. This forcing begins at $t=0$, and is applied to an initially stationary background. We ran two sets of simulations, with plane-wave forcing patterns having $[i,j,k] = [15,20,0]$ and $[25,5,0]$. We shall refer to these as `steep' and `shallow' (with respect to the stratification). These correspond to (the analogues of) axisymmetric modes as there is no $z$-($\varphi$-)dependence. The fluid in the forcing regions was driven at the frequency associated with the wavevector of the mode, as given by Equation (\ref{eq:gravity_DR}). These are $\omega_f = 0.8$ and 0.196 for the steep and shallow runs, where the subscript $f$ denotes `forcing'. To satisfy the incompressible condition, the fluid velocity $\mathbf{u}$ at any point $\mathbf{x}$ must obey $\mathbf{k} \cdot \mathbf{u} = 0$, i.e.~it must be proportional to the vector
\begin{align}
  \hat{\mathbf{u}} = \left( \kappa_\perp, -\frac{\kappa_x \kappa_y}{\kappa_\perp}, -\frac{\kappa_x \kappa_z}{\kappa_\perp} \right) \label{eq:forcing_u}
\end{align}
where $\kappa_x$, $\kappa_y$ and $\kappa_z$ are the components of $\mathbf{k}/|\mathbf{k}|$, and we additionally use the condition that the perpendicular components of $\mathbf{u}$ and $\mathbf{k}$ (with respect to the stratification) must be parallel. The functional form of the forcing function (fluid acceleration) is given by
\begin{align}
  \mathbf{f} \propto \begin{cases}
    \hat{\mathbf{u}} \cos (\mathbf{k} \cdot \mathbf{x}) \cos(\omega_f t) & \quad \text{if} \; |x| > 1.5 \\
    0 & \quad \text{otherwise,}
  \end{cases} \label{eq:forcing_fn}
\end{align}
with the overall amplitude for the moment arbitrary. In each run, internal gravity waves were generated and observed to propagate towards the centre of the box, whereupon a quasi-steady-state solution was eventually reached after several information crossing times. The amplitude of the forcing, i.e.~the scaling of Equation (\ref{eq:forcing_fn}), was set to be the same for all runs, and was such that maximum fluid velocities did not exceed $10^{-3}$. This is smaller than the group speed of gravity waves (0.015 and 0.025 for the steep and shallow runs), and translates to characteristic fluid displacements of 0.01 units, which is much smaller than the box width and extent of the field region. Note that the amplitudes of excited motions in solar-like oscillators are, likewise, much smaller than the stellar radius.

For each of the two sets of $[i,j,k]$, six different field strengths, both above and below the critical value, were simulated. In the non-dimensionalisation scheme used in \textsc{snoopy}, the critical field strength is given simply by $B/\sqrt{\rho} \sim \omega_f/|\mathbf{k}|$. This evaluates to 0.02 for the steep run and 0.005 for the shallow run. To control the strength of the field, we condensed the scalar multiplying factors in Equation (\ref{eq:B_soln}) into a single parameter, $B_s \equiv \kappa \rho/\lambda$. Note that the field strength varies over space, and so $B_s$ is only a characteristic value of the field rather than the maximum. The maximum field strength, attained at $R=0$, is about three times this. The choice of Reynolds numbers was constrained by the need for the length scale of physical (thermal and viscous) diffusion associated with the oscillation timescale to be smaller than the wavelength of the forced mode (0.16 units), and larger than the spacing of the grid (0.004 units). This sets a usable range between $\sim 10^4$--$10^6$. We opted for Re = Rt = $10^5$ for the steep run and Re = Rt = $4 \times 10^5$ for the shallow run. The four-fold difference in these values owes to a desire to maintain a similar diffusion lengthscale, since the associated timescales (wave periods) differ by a factor of about four. See Table \ref{tab:sims} for a summary of the parameters used.

\begin{table}
  \centering
  \caption{Parameters used for the twelve \textsc{snoopy} simulation runs. The box size of $[L_x,L_y,L_z] = [4,4,1]$ and resolution of $1024 \times 1024 \times 8$ are common between all runs, as are the location and extent of the magnetic field and forcing regions.}
  \begin{tabular}{ccccc}
    \hline
    $[i,j,k]$ & $\omega_f$ & $B_s$ & Re=Rt & Rm \\
    \hline
    \multirow{6}{*}{\makecell{$[15,20,0]$ \\ (steep)}} & \multirow{6}{*}{0.8} & 0 & \multirow{6}{*}{$10^5$} & \multirow{6}{*}{$10^6$} \\
    & & 0.005 & & \\
    & & 0.01 & & \\
    & & 0.02 & & \\
    & & 0.04 & & \\
    & & 0.08 & & \\
    \hline
    \multirow{6}{*}{\makecell{$[25,5,0]$ \\ (shallow)}} & \multirow{6}{*}{0.196} & 0 & \multirow{6}{*}{$4 \times 10^5$} & \multirow{6}{*}{$10^6$} \\
    & & $1.25 \times 10^{-3}$ & & \\
    & & $2.5 \times 10^{-3}$ & & \\
    & & 0.005 & & \\
    & & 0.01 & & \\
    & & 0.02 & & \\
    \hline
  \end{tabular}
  \label{tab:sims}
\end{table}

\begin{figure}
  \centering
  \includegraphics[width=\columnwidth]{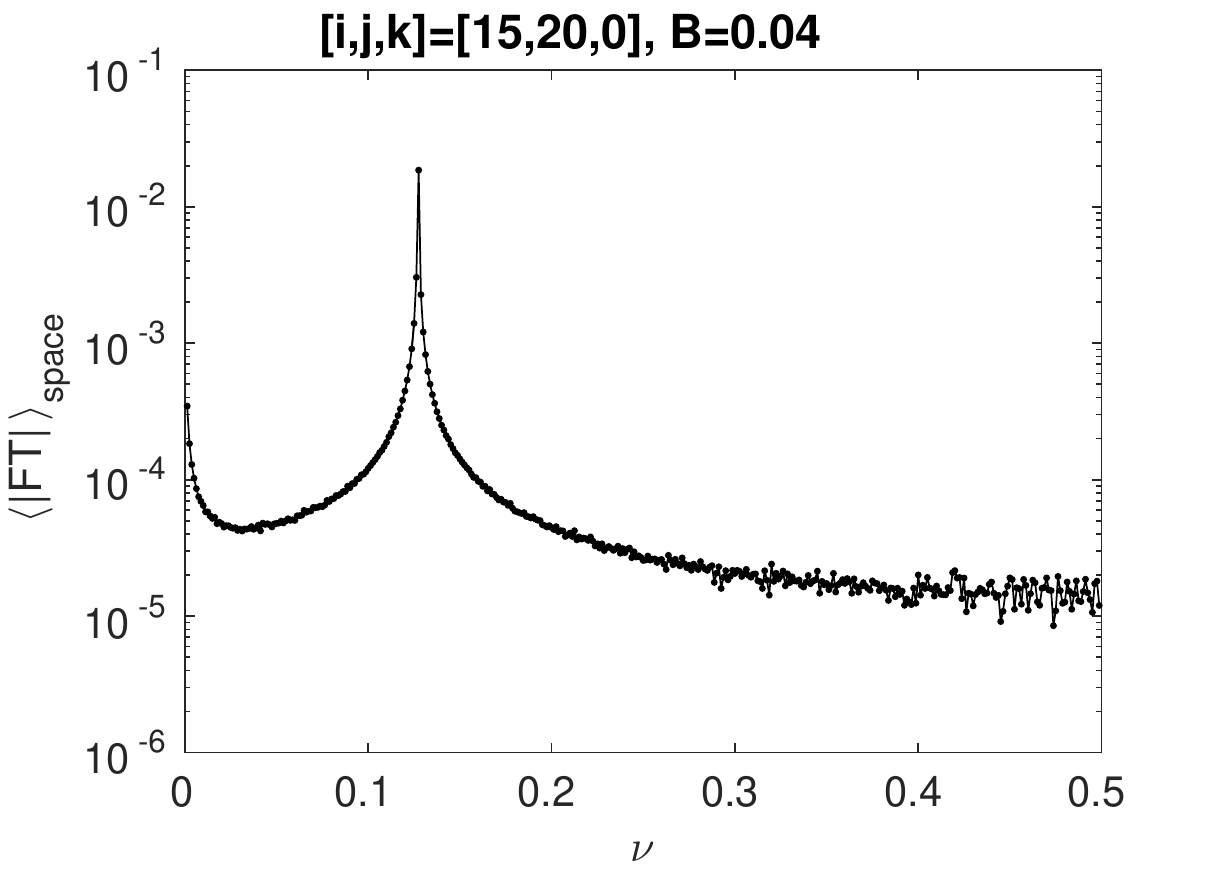}
  \caption{Spatially-averaged $v_y$ Fourier amplitude as a function of frequency, for the steep run, $B=0.04$, $200 < t < 1000$. The averaging has been restricted to the region within $|x| < 1.4$ to avoid the forcing strip. Note that the vertical axis is logarithmic. A substantial peak occurs at the value of the forcing frequency, $\nu = \omega/(2\pi) = 0.127$. Very similar plots are obtained for all other field strengths and also for the shallow run.}
  \label{fig:FTspace_vs_freq}
\end{figure}

\begin{figure}
  \centering
  \includegraphics[width=\columnwidth]{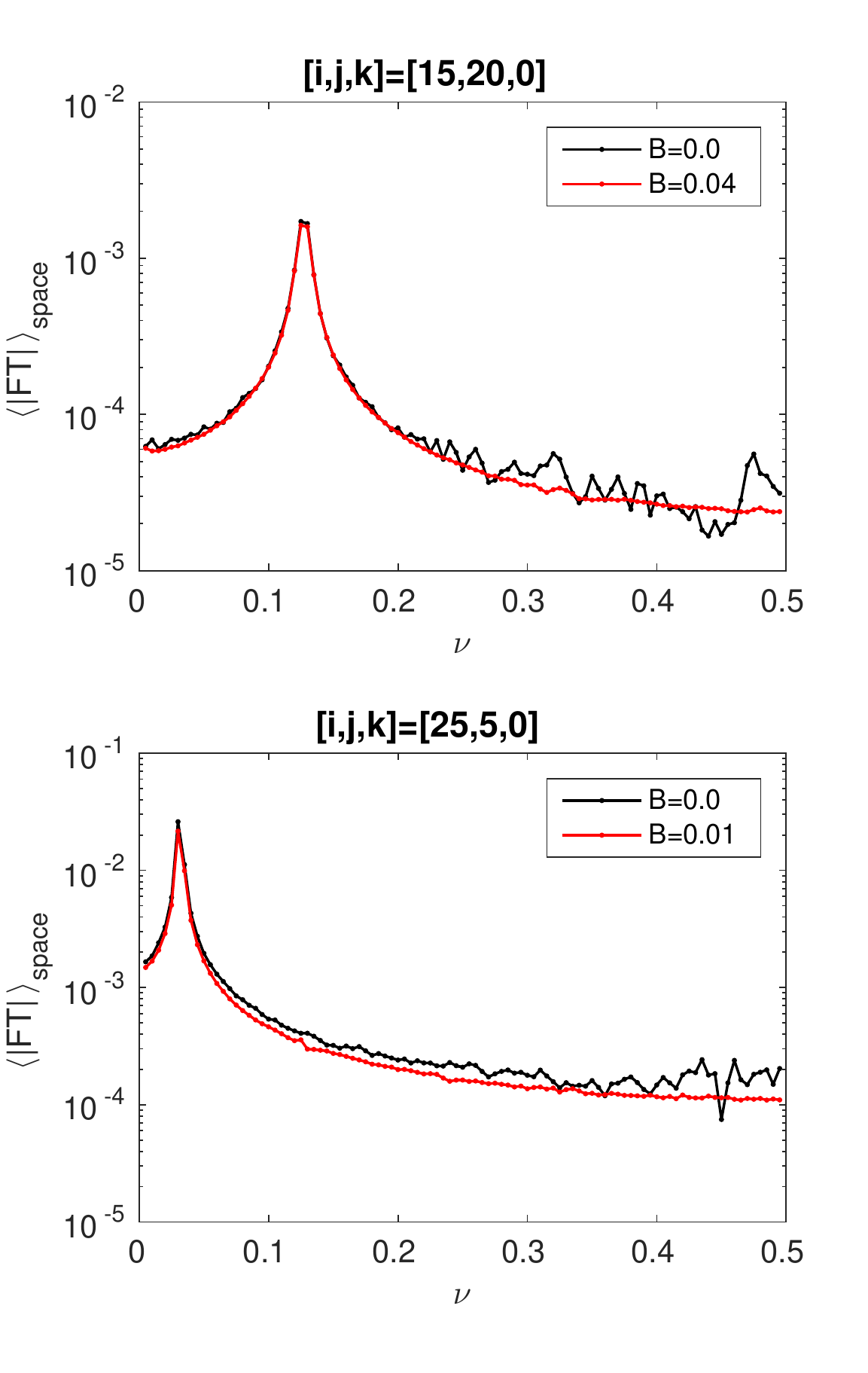}
  \caption{Spatially-averaged $v_y$ Fourier amplitude as a function of frequency, computed over the interval $1000 < t < 1200$, i.e.~after the forcing has been turned off, for the steep (top) and shallow (bottom) runs. The averaging is taken over the whole box. Black corresponds to purely hydrodynamic runs, while red corresponds to strong-field runs. Even though the systems are no longer forced, peaks are still seen at the locations of the original forcing frequencies in all cases, suggesting that these are natural response frequencies.}
  \label{fig:FTspace_vs_freq_forceoff}
\end{figure}

\begin{figure}
  \centering
  \includegraphics[width=\columnwidth]{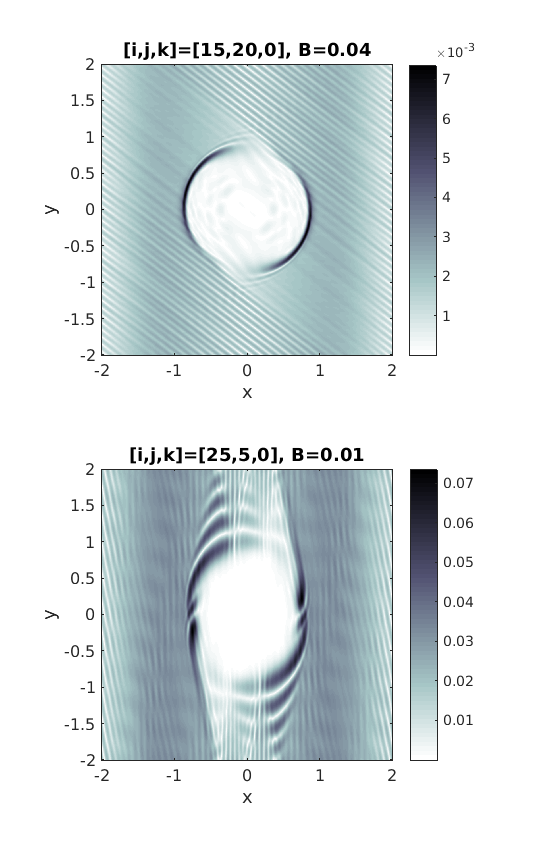}
  \caption{Spatial distribution of the $v_y$ Fourier amplitudes computed over the interval $1000 < t < 1200$ (forcing switched off), at the frequencies corresponding to the peaks in Figure \ref{fig:FTspace_vs_freq_forceoff}. Note the qualitative similarity to their forced counterparts in Figures \ref{fig:steep_vy_tFT} and \ref{fig:shallow_vy_tFT}.}
  \label{fig:vy_tFT_force_off}
\end{figure}

The characteristic box-crossing times (in the $x$-direction, based on the gravity-wave group velocity) are 80 and 200 time units for the steep and shallow runs, respectively. Each simulation was allowed to proceed for 1200.0 time units, with the velocity fields sampled at intervals of 1.0 time units. For comparison, forcing periods were 7.85 and 32.1 units for the steep and shallow runs, consistent with Equation (\ref{eq:gravity_DR}). We chose to discard the first 200 time units of data in subsequent analysis, to remove transient signals associated with the settling-in phase. To investigate the natural response of the system, the forcing was switched off at $t=1000.0$. Note that plots of the results in subsequent sections are restricted to either the forced interval ($200.0 < t < 1000.0$) or the unforced interval ($1000.0 < t < 1200.0$). The periodic nature of the solutions in space and time suggests a convenient analysis approach to be temporal and spatial Fourier transforms, such as shown in Figures \ref{fig:steep_vy_tFT} and \ref{fig:shallow_vy_tFT}.

As discussed in \S\ref{sec:cylPrendergast}, the analogous coupling of spheroidal and torsional motions in our cylindrical geometry does not naturally occur, and so to investigate this we modified the functional form of the Lorentz force within \textsc{snoopy}. To artifically couple $x,y$ motions into $z$, we added the equivalents of the curvature term (proportional to $1/r$) to the $z$-component of the Lorentz force. More specifically, we defined a new parameter $r_\text{fake}$ and replaced
\begin{align}
  \mathbf{B} \cdot \nabla B_z \to \mathbf{B} \cdot \nabla B_z + \frac{B_x B_z - B_{0x} B_{0z}}{r_\text{fake}} \:, \label{eq:hack}
\end{align}
where $B_{0x}$ and $B_{0z}$ are the $x$ and $z$ components of the initial field configuration given by Equation (\ref{eq:B_soln}), and the remaining terms correspond to values at the current timestep in the simulation. The coupling parameter $r_\text{fake}$ represents the characteristic radius of curvature in the symmetry direction. For the spherical Prendergast solution, this is of order the size of the field region itself, and so we set $r_\text{fake} = a$.

Compared to the work of \citet{Lecoanet2017}, our setup differs in several key ways. Firstly, we include both a substantial zero-field region as well as a region containing a field, with forcing applied to the fluid in the zero-field region (thus sending waves into a magnetised region from outside). In contrast, the setup of \citet{Lecoanet2017} has no regions within the domain that are completely devoid of field. Their forcing is applied directly to a part of the magnetised region, above which the maximum field strength over the domain rises by no more than a factor of three. This means that the scattering problem is not well-defined: such a setup by its confined nature is likely to exclude the possibility of observing wave reflection. Secondly, their configuration has magnetic field lines rooted in artificial damping regions on the boundaries of the domain, preventing the formation of standing vibrations on closed field loops, which potentially play a role in the dynamics. In contrast, our cylindrical Prendergast configuration derived in \S\ref{sec:cylPrendergast} describes a magnetic torus, with all field lines closing within the simulation domain. It is also to be noted that the resolution of their setup is organised to represent small-scale disturbances with wavevectors almost parallel to the stratification, which may limit the possible range of outcomes. Our simulations do not suffer this limitation, as they are well resolved both parallel and perpendicular to the stratification. For these reasons, our setup, though not fully global, better represents an actual stellar interior, and could explain why a greater variety/complexity of phenomena (see later in \S\ref{sec:results}) are observed compared to their previous work.

\begin{figure*}
  \centering
  \includegraphics[clip=true, trim=2cm 0cm 2cm 0cm, width=\textwidth]{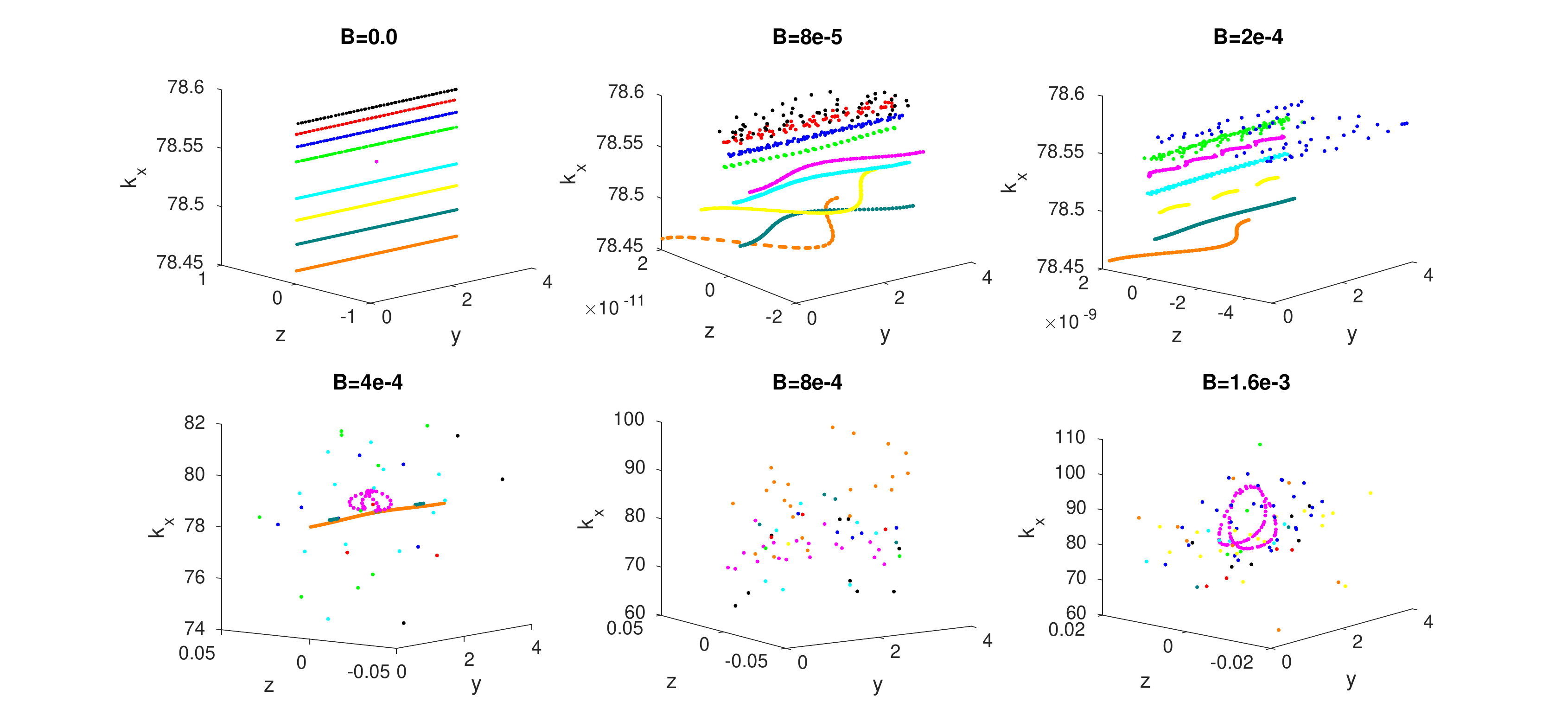}
  \caption{Poincar\'{e} surfaces of section (see \S\ref{sec:MG_ray_tracing} for an explanation) for a collection of nine rays (different colours) with common $|\mathbf{k}|$, launched at slightly different angles about the ray with $[i,j,k] = [50,2,0]$ from $(x,y,z)=(1,-1,0)$. This corresponds to a shallow incoming angle (N.~B.~shallower than for the shallow \textsc{snoopy} run). The six panels are for different field strengths. Only for the lower three panels do critical surfaces exist. Note that the plotting limits have been truncated and adjusted for clarity, meaning that some points (such as the red and black rays in the top-right panel, which are sparsely scattered about a larger 3D volume) are not shown.}
  \label{fig:PSS_i50_j2_k0}
\end{figure*}

\subsection{Magneto-gravity ray tracing}\label{sec:MG_ray_tracing}
The phase-space evolution of ray trajectories is governed by a Hamiltonian system and is given by solving
\begin{align}
  \frac{\rmd \mathbf{x}}{\rmd t} = \nabla_\mathbf{k} H \:, \quad
  \frac{\rmd \mathbf{k}}{\rmd t} = -\nabla H \:, \label{eq:Hamilton}
\end{align}
where $t$ is time, $\mathbf{x}$ are the spatial coordinates, $\mathbf{k}$ are the conjugate momenta, and $H = H(\mathbf{x}, \mathbf{k}, t)$ is the Hamiltonian. While commonly encountered in the context of particle trajectories in solid mechanics, Equations (\ref{eq:Hamilton}) are also applicable to wave phenomena in continuum mechanics, having for example been used extensively in the study of MHD wave propagation in the Earth's magnetosphere \citep{Haselgrove1955, Dyson1967, Walker2004, Fung2005}. In the wave context, the wave frequency given by the dispersion relation $\omega = \omega(\mathbf{x}, \mathbf{k}, t)$ takes on the role of the Hamiltonian and Equations (\ref{eq:Hamilton}) trace the group-velocity paths of the rays, i.e.~the trajectories of wave packets with wavenumber $\mathbf{k}$. This corresponds to the WKB limit of the usual fluid equations, analogous to the geometric optics limit of electromagnetism or the classical limit of quantum mechanics. It is formally exact in the limit of short wavelengths, a generally good approximation for solar-like oscillators in which the length scales of disturbances are much smaller than those of background variations. There are several advantages of this approach over full-wave calculations, namely the computational speed and simplicity of integrating a system of linear ODEs, and the ability to handle arbitrarily complex, fully three-dimensional background geometries at no significant additional computational expense. Although at the level of approximation used here it is unable to treat partial wave packet reflection/conversion, it nonetheless remains a powerful technique for visualising the dominant paths of energy flow and provides useful physical insights into the results of numerical simulations \citep[e.g.][]{Alvan2015}.

In deep stellar interiors, sound speeds greatly exceed those associated with other wave modes, and so their role in the dynamics can be neglected. A WKB treatment of the fluid equations, ignoring acoustic effects, leads to the dispersion relation for magneto-gravity waves \citep{Unno1989}
\begin{align}
  \omega^2 = \omega_A^2 + \kappa_\perp^2 N^2 \:, \label{eq:MG_DR}
\end{align}
where $\omega_A \equiv \mathbf{k} \cdot \mathbf{v}_A$ is the Alfv\'{e}n frequency, and $\mathbf{v}_A$ is the Alfv\'{e}n velocity. Setting $H(\mathbf{x}, \mathbf{k}, t) = \omega(\mathbf{x}, \mathbf{k}, t)$, and taking into account $z$-translational symmetry, Equations (\ref{eq:Hamilton}) with respect to Cartesian coordinates $(x,y,z)$ become
\begin{align}
  \frac{\rmd \mathbf{k}}{\rmd t} &= -\frac{\omega_A}{\omega} \left( \mathbf{k} \cdot \frac{\del \mathbf{v}_A}{\del x} \:,\; \mathbf{k} \cdot \frac{\del \mathbf{v}_A}{\del y} \:,\; 0 \right) \:, \label{eq:ray_tracing_1} \\
  \frac{\rmd \mathbf{x}}{\rmd t} &= \frac{\omega_A}{\omega} \mathbf{v}_A + \frac{N^2 \kappa_x}{|\mathbf{k}| \omega} \left( -\kappa_\perp^2 \:,\; \kappa_x \kappa_y \:,\; \kappa_x \kappa_z \right) \:. \label{eq:ray_tracing_2}
\end{align}
For the cylindrical Prendergast solution,
\begin{align}
  \frac{\del \mathbf{v}_A}{\del x} &= \frac{B_s}{J_0(\lambda a) R} \left( xy f(R) \:,\; -x^2 f(R) - J_1(\lambda R) \:,\; \lambda x J_1(\lambda R) \right) \:, \label{eq:dvAdx} \\
  \frac{\del \mathbf{v}_A}{\del y} &= \frac{B_s}{J_0(\lambda a) R} \left( y^2 f(R) + J_1(\lambda R) \:,\; -xy f(R) \:,\; \lambda y J_1(\lambda R) \right) \:, \label{eq:dvAdy} \\
  \shortintertext{where}
  f(R) &\equiv \frac{\lambda}{R} J_0(\lambda R) - \frac{2}{R^2} J_1(\lambda R) \:, \label{eq:fR_defn} \\
  R &\equiv \sqrt{x^2 + y^2} \:. \label{eq:R_defn}
\end{align}

See that $\omega$ has no explicit time-dependence, and so represents a conserved quantity of motion. Equations (\ref{eq:ray_tracing_1})--(\ref{eq:ray_tracing_2}) were integrated using an explicit fourth-order Runga-Kutta scheme with a time step of 0.05 units up to a maximum time of $10^6$. Note that the scheme does not guarantee conservation of $\omega$; rather, we used the time variation of $\omega$ as an independent accuracy check. We verified that global fluctuations in $\omega$ were typically less than one part in $10^6$. A constant buoyancy frequency of $N^2 = 1$ with stratification in the $x$-direction, just as for the \textsc{snoopy} simulations, was used in all ray-tracing calculations.

With three spatial degrees of freedom, the phase space in the absence of any symmetries is six-dimensional: a given point is defined by three spatial coordinates $(x,y,z)$ and three conjugate momenta $(k_x,k_y,k_z)$. If one were to choose an initial set of $(k_x,k_y,k_z,x,y,z)$ and integrate Equations (\ref{eq:ray_tracing_1})--(\ref{eq:ray_tracing_2}), the resulting trajectory would wander around a six-dimensional volume. In our problem, however, two constants of motion exist, which confine trajectories to a four-dimensional surface. This includes the Hamiltonian itself (since there is no explicit time-dependence), and also $k_z$ due to $z$-translational symmetry. We make use of Poincar\'{e} surfaces of section (PSS) to analyse the phase-space behaviour of the rays \citep[e.g.][]{Dittrich2001}. This technique involves artificially fixing one of the free parameters and plotting the remaining ones against each other. With four free parameters in our case, the PSS is three-dimensional. In all phase-space plots presented here, we have fixed the $x$-coordinate and plotted $k_x$ as a function of $y$ and $z$ (see Figure \ref{fig:PSS_i50_j2_k0}). For a given ray trajectory, the values of $(y, z, k_x)$ are recorded each time the ray crosses the surface $x=2$ and plotted as a point on a three-dimensional scatter plot.

\begin{figure*}
  \centering
  \includegraphics[width=0.9\textwidth]{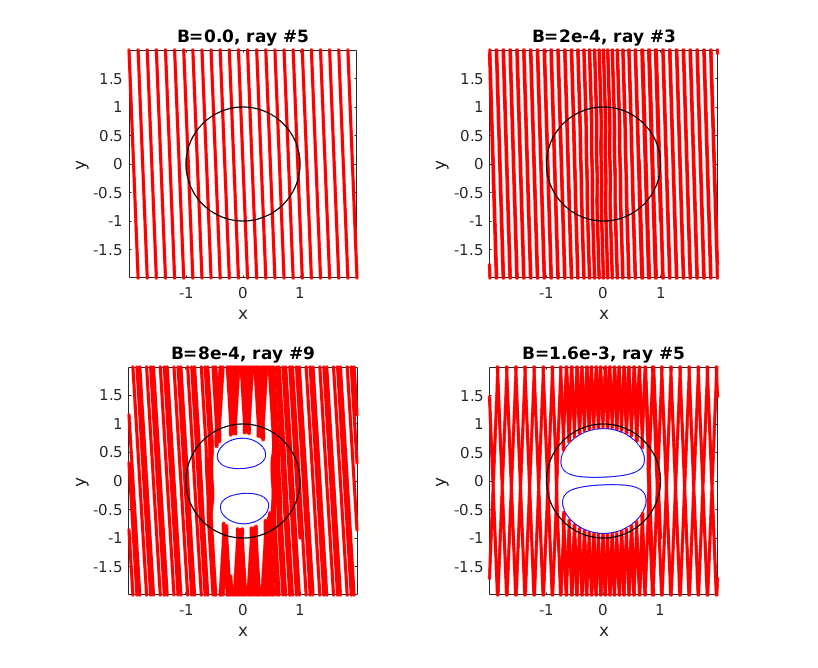}
  \caption{Trajectories in space for a selection of rays from Fig.~\ref{fig:PSS_i50_j2_k0}, overlaid with the field region boundary (black) and critical surfaces associated with the initial wavevector of the ray (blue). Rays are numbered such that \#5 refers to magenta, \#3 to blue and \#9 to orange in Fig.~\ref{fig:PSS_i50_j2_k0}. Note that ray paths in general move in three spatial dimensions, but only the $x,y$-projection is shown for simplicity. The trajectories shown were computed up to $t = 2.5 \times 10^5$.}
  \label{fig:trajectories}
\end{figure*}

\begin{figure}
  \centering
  \includegraphics[width=\columnwidth]{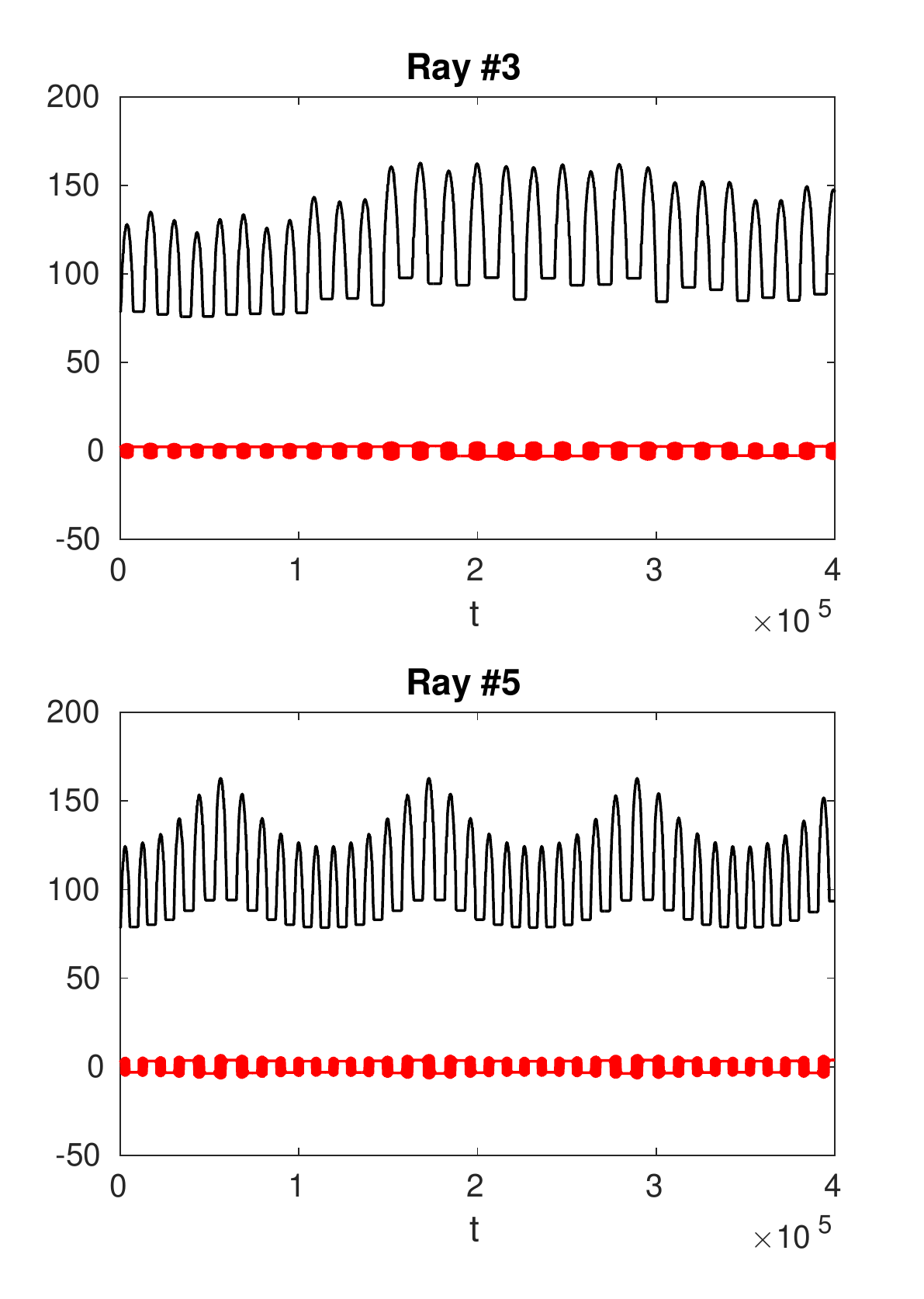}
  \caption{Evolution of $k_x$ (black) and $k_y$ (red) as a function of time, for the blue (top) and magenta (bottom) rays of the bottom-right panel in Fig.~\ref{fig:PSS_i50_j2_k0} ($B = 1.6 \times 10^{-3}$).}
  \label{fig:kcompts_chaotic_stable}
\end{figure}

\begin{figure}
  \centering
  \includegraphics[width=\columnwidth]{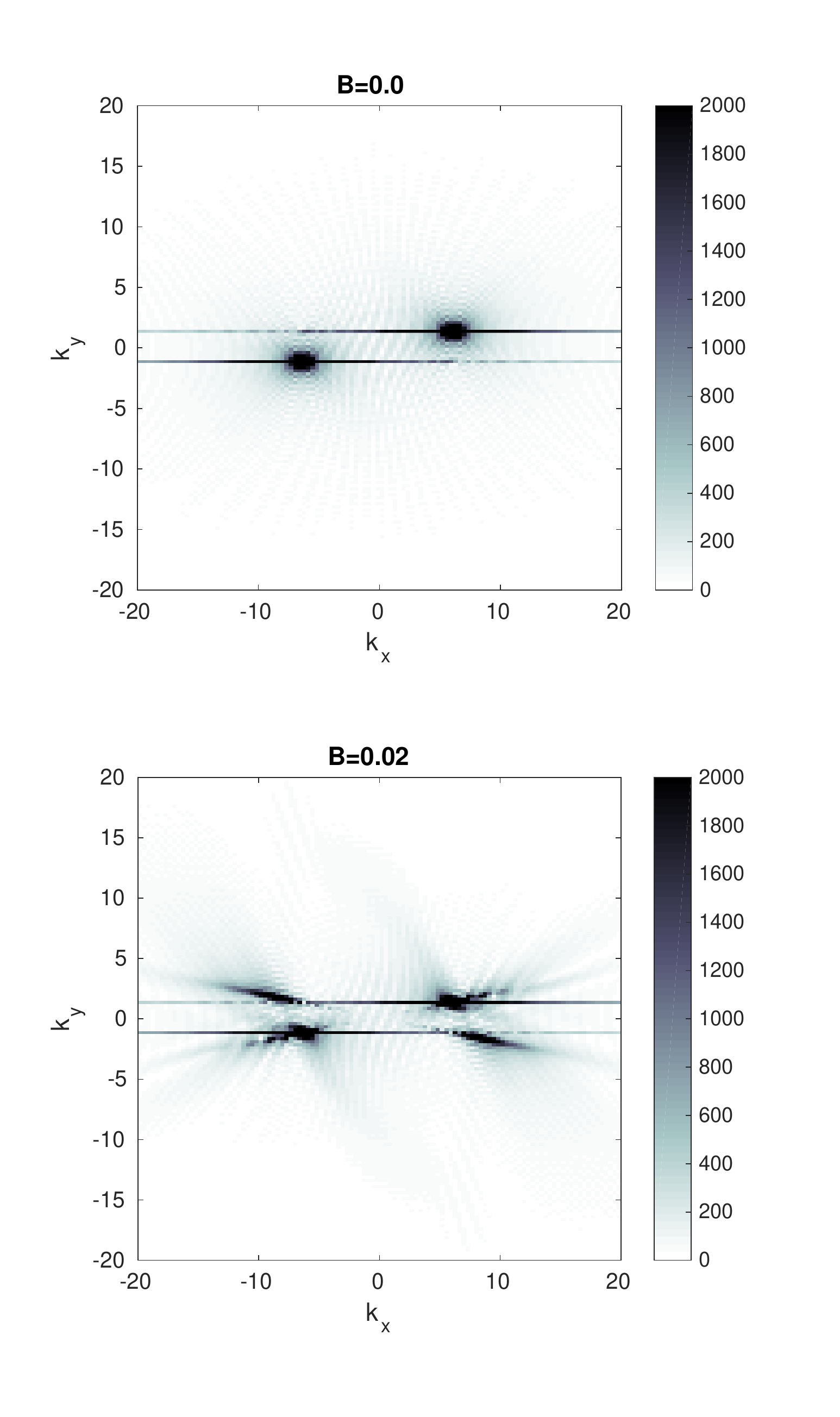}
  \caption{Two-dimensional spatial Fourier transform of the $v_y$ field at the forcing frequency for $[i,j,k]=[25,5,0]$, taken over the region exterior to the field and interior to $|x| < 1.4$ (thus avoiding the forcing strip, which occupies $|x| > 1.5$). The temporal transform is taken over the forced interval, $200 < t < 1000$. The top panel corresponds to the zero-field case, and shows peaks at the forcing wavenumber. The bottom panel is for the strongest field strength simulated, and shows significant amounts of power at both the forcing wavenumber and approximately its negation, evidencing specular reflection.}
  \label{fig:vy_sFT_i25_j5_k0}
\end{figure}

\section{Results}\label{sec:results}

\subsection{Modification to g-mode structure}
Figures \ref{fig:steep_vy_tFT} and \ref{fig:shallow_vy_tFT} show the amplitudes of the temporal Fourier transform of the $y$-component of the fluid velocity, at each point on the grid, at the value of the forcing frequency (computed over the forced interval $200.0 < t < 1000.0$) in the \textsc{snoopy} simulations. Very similar patterns are seen in $v_x$ (since this is related to $v_y$ through $\nabla \cdot \mathbf{v} = 0$), and so the corresponding plots are not shown. Although the simulations are non-linear, it is still the case that the box-averaged absolute fluid velocity peaks sharply at the forcing frequency (see Figure \ref{fig:FTspace_vs_freq}), and so we have chosen this value of $\omega$ for displaying the results. The six panels show different values of the field strength going from zero (purely hydrodynamic) to strongly magnetised. A convenient indication of the strength of the field is the presence or absence of what we shall refer to as \textit{critical surfaces}, i.e.~where the Alfv\'{e}n frequency at the forced wavenumber equals the forcing frequency. These surfaces are plotted in blue. In the absence of modifications to the wavelength of the incoming waves, they physically represent surfaces where the resonance criterion is satisfied and the magnetic field might be expected to become important. We shall loosely refer to the simulations in which no critical surfaces exist as `weak-field' and those in which they do as `strong-field'.

In both steep and shallow sets of simulations, a qualitative transition in wave propagation coincides with the transition from weak to strong fields (appearance of critical surfaces). In the weak-field regime, small distortions to the gravity wave phase fronts are present within the field region, but the fronts otherwise remain intact. In the strong-field regime these become severely disrupted, but the way in which this happens depends crucially on their orientation. For the steep runs, the wave energy at the point of disruption becomes concentrated into arcs aligned along the flux surfaces roughly tangent to the critical surfaces. There is little penetration of fluid motion past this point. The efficiency of the expulsion increases with field strength (note that the same colour scale has been used on all panels of the same figure). For the shallow runs, rather than being trapped within the field region, there is instead reflection of the waves off the critical surfaces and back out of the field region. This can be seen through the organised interference patterns that form above and below the cylinder; note that the group velocity in the shallow case is directed nearly parallel to the $y$-axis, and so the location of the interference patterns supports the interpretation that the reflection is of a near-specular kind with respect to the stratification, where only the $y$-component of the group velocity flips sign.

\begin{figure}
  \centering
  \includegraphics[width=\columnwidth]{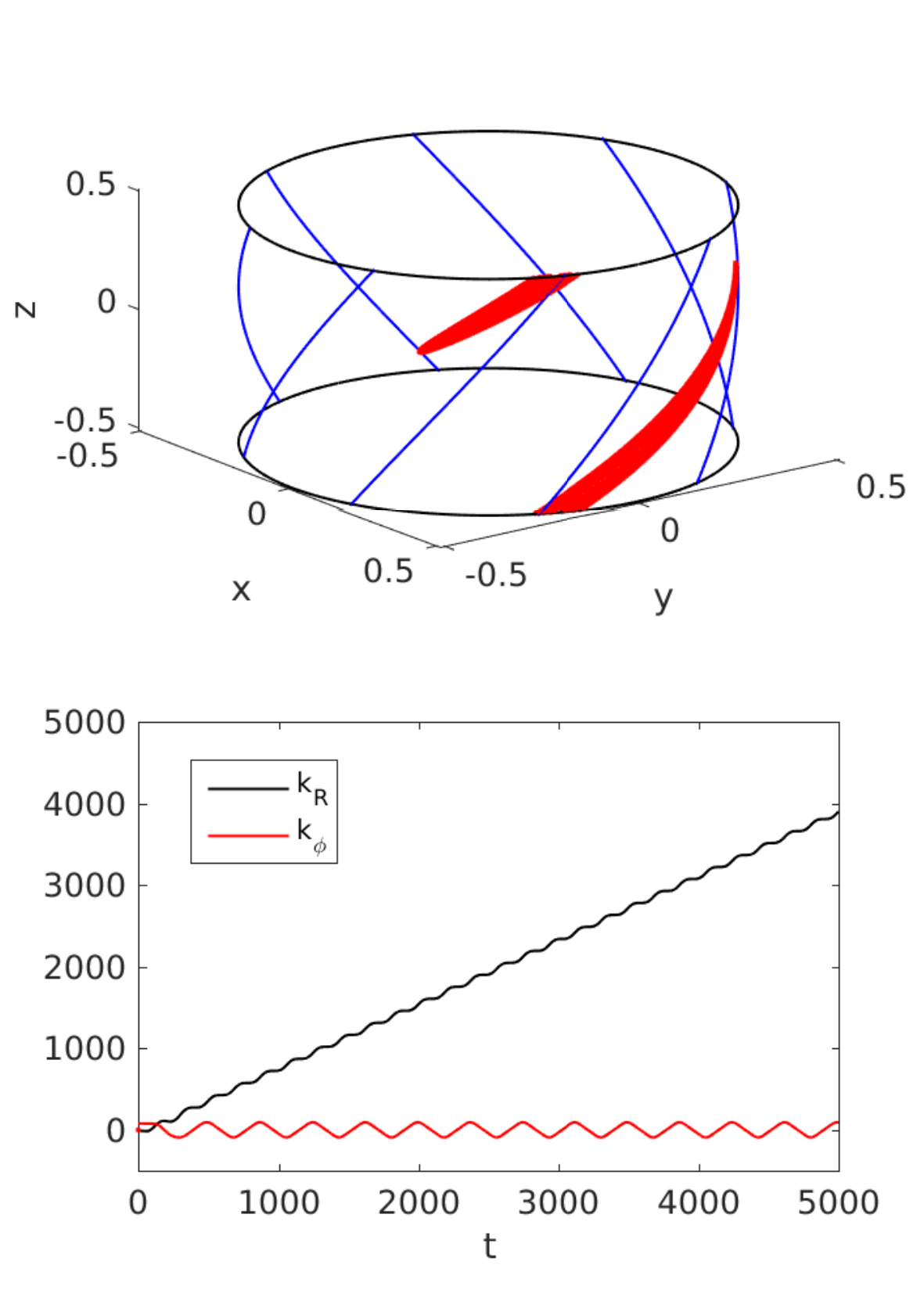}
  \caption{Top: Trajectory of a ray with initial $[i,j,k]=[30,40,0]$, restricted to $1000 < t < 5000$ (red). The path appears to be confined to the flux surface at radius $R \approx 0.51$. Black circles are a visual aid to marking the location of this flux surface. In blue are a selection of magnetic field lines on that surface. Note that periodic boundary conditions have been used for the ray tracing, so that the two apparently disjoint sections are in fact a continuous path. Bottom: The $R$ and $\phi$-components of the wavevector as a function of time. It can be seen that on average, $k_R$ undergoes linear growth while $k_\phi$ remains constant, a property expected of Alfv\'{e}n wave phase mixing.}
  \label{fig:trapped_ray}
\end{figure}

\begin{figure}
  \centering
  \includegraphics[width=\columnwidth]{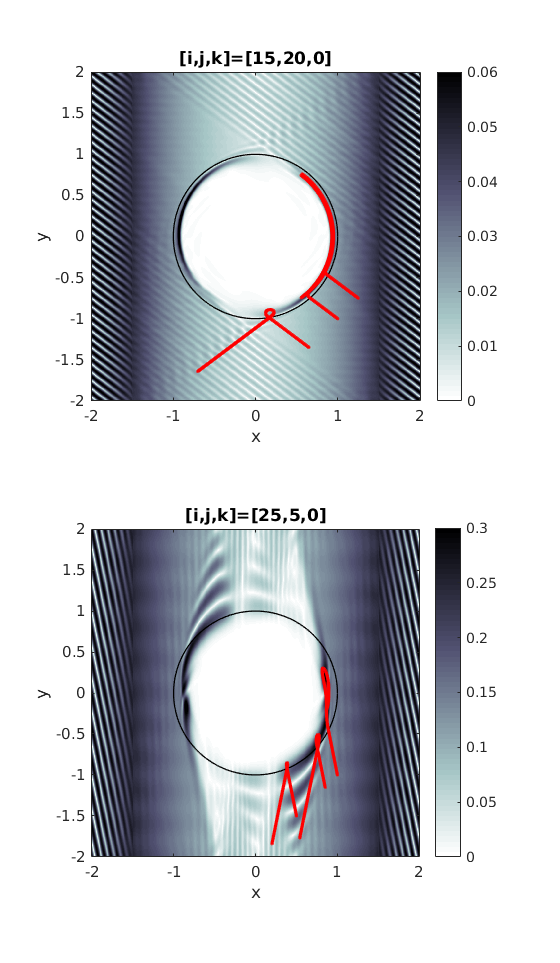}
  \caption{Spatial distribution of the $v_y$ temporal Fourier amplitudes at the respective forcing frequencies for $[i,j,k]=[15,20,0]$ (top) and $[i,j,k]=[25,5,0]$ (bottom). The temporal transform is taken over the forced interval ($200 < t < 1000$). In both cases the highest field strength is being shown ($B=0.08$ and $B=0.02$, respectively). Overlaid for comparison in red are the trajectories of three rays launched from along the line $y=x-2$, with the same initial wavenumber as the forcing function. In the top panel the two rays with launch points furthest on the right are trapped while the leftmost one is reflected, and in the bottom panel the two rays launched furthest on the left are reflected while the rightmost one is trapped. This behaviour closely matches the reflection/interference patterns and high-amplitude arcs shown in greyscale (the results of \textsc{snoopy} simulations).}
  \label{fig:vy_tFT_w_rays}
\end{figure}

\subsection{Survival of modes in strong-field regime}\label{sec:survival}
The forcing frequencies of $\omega_f = 0.196$ and 0.8 for the shallow and steep runs were chosen to correspond to those of pure gravity modes. We saw in the previous section that during the interval where the forcing is continuously applied, the response of the system peaks at the corresponding frequency. This is an expected property of any forced oscillator, whether or not the forcing frequency matches any natural frequency of the system. One might be curious as to whether the presence of a strong magnetic field destroys the ability of the system to support normal modes, as previously suggested by other authors \citep{Fuller2015, Lecoanet2017}. In order to test this, we turned the forcing off at $t = 1000.0$ and allowed the system to evolve naturally. We considered both steep and shallow runs, at two values of the field strength: zero, and the second-largest value.

If the steady-state patterns in Figures \ref{fig:steep_vy_tFT} and \ref{fig:shallow_vy_tFT} do not in fact correspond to normal modes, one expects their spatial and temporal coherence to break down within several oscillation periods after the forcing ceases. On the contrary, we find that the velocity field continues to undergo periodic oscillations at a frequency equal to that of the forcing (within measurement precision). Furthermore, there is no change to the spatial structure of the pattern associated with this frequency. On top of these oscillations, and independent of the field strength, there is a slower exponential decay that can be attributed to viscous dissipation. Figure \ref{fig:FTspace_vs_freq_forceoff} shows the spatially-averaged amplitudes of the temporal Fourier transform of the $v_y$ fields for each of the four cases (steep and shallow, zero field and strong field). A clear peak appears in all cases, at the frequency matching the original forcing frequency for that run. This frequency is the same regardless of whether or not there is a field present, which suggests that a strong field does not disrupt the temporal coherence of the oscillation. Figure \ref{fig:vy_tFT_force_off} shows the spatial distribution of the Fourier amplitudes associated with the frequency peaks in Figure \ref{fig:FTspace_vs_freq_forceoff}. They are remarkably similar to the patterns seen in Figures \ref{fig:steep_vy_tFT} and \ref{fig:shallow_vy_tFT}, suggesting that these do in fact correspond to normal modes of oscillation in the magnetised case. Importantly, these results demonstrate that coherent oscillations can be sustained in the presence of a strong magnetic field, even if the spatial structure is significantly modified compared to the hydrodynamic case, and the associated frequencies may not be significantly different from the pure g-modes.

\subsection{Transition to dynamical chaos}
A complementary approach to understanding the dynamics is through Hamiltonian ray tracing, which offers a means of quantifying the phase-space evolution of a wave packet propagated through the system. Compared to full non-linear wave calculations, ray tracing suffers fewer practical restrictions on the range of input parameters and is computationally inexpensive to implement. This allowed us to experiment with a wider range of input wavevectors compared to the two values used in the \textsc{snoopy} calculations, which were $[i,j,k]=[15,20,0]$ and $[25,5,0]$. Besides computing ray trajectories for the two above values, which afford direct comparison, we also computed these in addition for $[i,j,k]=[50,10,0]$, $[50,5,0]$, $[50,2,0]$, $[30,40,0]$ and $[45,25,0]$. For each $[i,j,k]$ combination, we tested a range of field strengths going from zero to above the critical value, just as for the \textsc{snoopy} calculations.

In all cases, the weak-strong magnetic field transition is associated with a transition in the ray dynamics from regular to chaotic. This can be seen in Figure \ref{fig:PSS_i50_j2_k0}, which plots the resulting PSS for $[i,j,k]=[50,2,0]$. The phase-space trajectories of a set of nine rays launched from the same closely-spaced initial positions progressively diverge as the field strength increases. At zero field strength (top left panel) all trajectories on the PSS are one-dimensional curves, reflecting the existence of four conserved quantities ($\omega, k_x, k_y$ and $k_z$ as discussed previously). These curves deform as the field increases away from zero. At low field strengths, many retain their one-dimensional (regular) nature, but this gives way to the emergence of multi-dimensional structures at higher field strengths. As can be seen in the middle and right panels along the top row, there is a spreading of some trajectories first into two-dimensional sheet structures, and then three-dimensional volumes (bottom row). Amidst these large, chaotic regions of phase space, there still exist a small number of regular modes (e.g.~magenta trajectory in the bottom-right panel) even at high field strengths. However, these are comparatively rare; it is likely that in the strong-field regime most of the modes will be chaotic.

Figure \ref{fig:trajectories} shows the spatial paths of a selection of regular and chaotic rays. Note that for the sake of clarity the plotting has been restricted to the first quarter of the total integration. In the top-right and bottom-left panels the rays, both of which are chaotic, eventually fill a large fraction of the $x,y$-plane (excluding critical regions, which reflect the rays). The remaining two (top-left and bottom-right panels) are regular and fill a smaller fraction of the $x,y$-plane (the ray paths loop back upon themselves after a finite amount of time). Ergodicity is a general property of chaotic modes, and so it would be na\"{i}vely expected that the associated regions of the PSS should eventually be completely filled with points. However, despite integrating for a large number of box-crossing times, the bottom row of Figure \ref{fig:PSS_i50_j2_k0} remains sparsely filled. The reason for this is rather curious: it appears to be a consequence of a trapping process associated with the existence of the critical surfaces. Rays which happen to impinge upon the field region in certain locations are thereafter trapped on quasi-periodic bounce orbits, never to re-emerge. This phenomenon is discussed in further detail in the following section.

\begin{figure*}
  \centering
  \includegraphics[clip=true, trim=2cm 0cm 2cm 0cm, width=\textwidth]{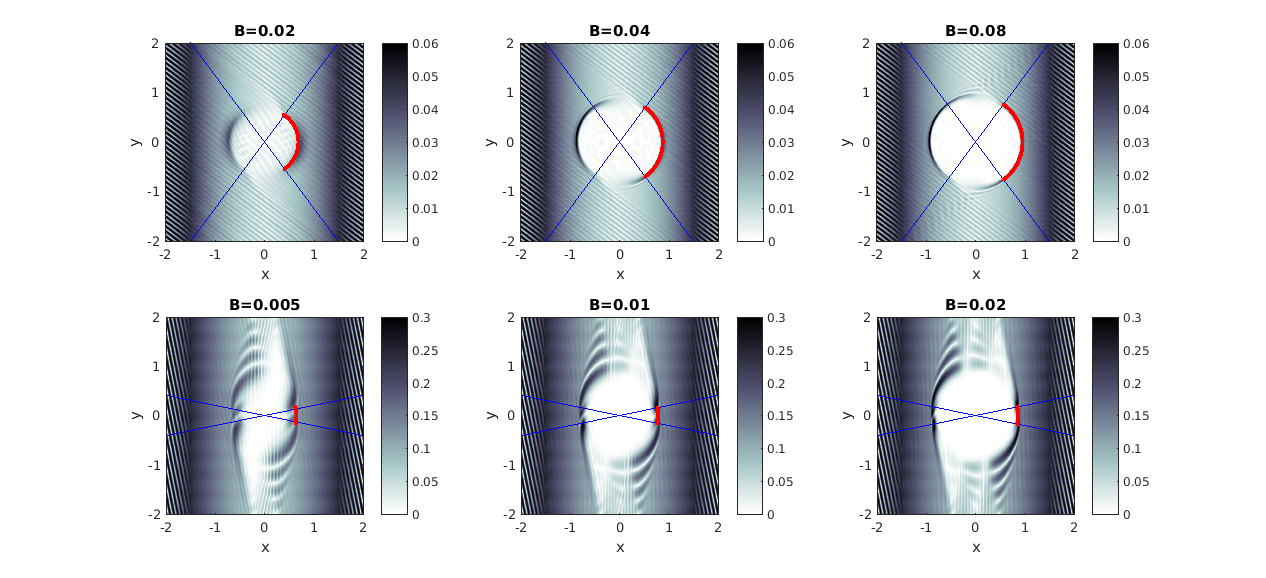}
  \caption{Spatial distribution of the $v_y$ temporal Fourier amplitudes at the respective forcing frequencies for $[i,j,k]=[15,20,0]$ (top row) and $[i,j,k]=[25,5,0]$ (bottom row), for the three strongest field strengths simulated in each case. Overlaid in blue are lines subtending the trapping angle $\phi_*$ predicted by Equation (\ref{eq:trapping_angle}), and in red the path of a ray launched into the field region that subsequently becomes trapped (only the part of the trajectory after trapping occurs has been plotted). There is excellent agreement between the ray trajectories and the analytical prediction, and a fair correspondence with features in the \textsc{snoopy} simulations.}
  \label{fig:trapping_angles}
\end{figure*}

\subsection{Wave trapping and reflection}\label{sec:trapping_reflection}
The two lower panels of Figure \ref{fig:trajectories} demonstrate the reflection of rays off the magnetic field region, specifically in the vicinity of the critical surfaces, which are marked in blue. See that the ingoing and outgoing angles with respect to the stratification direction are the same: this must be the case if we are to conserve $\omega$ as per Equation \ref{eq:gravity_DR}. However, the dispersion relation alone sets no constraint on the magnitude of the outgoing wavevector with respect to the ingoing value. Previous work based solely on heuristic arguments \citep{Fuller2015} suggested that upon reflection, the waves would be upscattered to systematically higher wavenumbers (the ``magnetic greenhouse effect''). Based on our \textsc{snoopy} simulations and Hamiltonian ray calculations, we find this not to be the case.

Figure \ref{fig:kcompts_chaotic_stable} shows the time evolution of the wavevector components of two rays, one chaotic and one regular, from the $[i,j,k]=[50,2,0]$, $B=1.6\times 10^{-3}$ ray tracing run. While there are certainly fluctuations in the magnitude of the wavevector, no systematic growth is seen over the course of many reflections. Figure \ref{fig:vy_sFT_i25_j5_k0} illustrates the same quantities measured from the \textsc{snoopy} simulations for the shallow run, where the wavevector components have been obtained by taking a spatial Fourier transform of the $y$-component of the velocity field. The transformed region excludes the magnetic cylinder and the forcing strip. In the field-free case (top panel), peaks are only seen at the forced wavevector, as per expectation. In the presence of a strong field (bottom panel), we see that in addition to the peaks at the forced wavevector there is a second pair of peaks at roughly its negation. Although a small amount of power trails off to higher wavenumbers, the response is still dominated by wavevectors with magnitudes similar to the input value. This suggests that the magnetic greenhouse effect may not be important, if it occurs at all. It also foreshadows the possibility of normal modes in the strongly magnetised regime, since the organised manner of reflection (as opposed to a random scattering) might enable structured interference patterns to form and persist, such as those found in \S\ref{sec:survival}.

Besides reflection, a second distinct phenomenon is noted to occur, and that is the trapping of rays within the field region itself. Examples of these are shown in close-up view in Figure \ref{fig:trapped_ray} (top) and in projected view along the $z$-axis in Figure \ref{fig:vy_tFT_w_rays}. From a ray-tracing perspective, a ray launched into the field region from outside appears to propagate inward until it reaches a flux surface roughly coincident with the critical surface, whereupon its trajectory veers to become roughly tangential to the flux surface. Physically, this can be interpreted as a conversion to some kind of modified Alfv\'{e}n wave (note that the trajectories do not follow field lines exactly, and so these are not pure Alfv\'{e}n waves). The bottom panel of Figure \ref{fig:trapped_ray} plots the time evolution of the wavevector components, decomposed locally into the cylindrical polar $R$ and $\phi$ directions. It can be seen that the radial component of the wavevector grows roughly linearly with time. Mathematically, this can be understood in the framework of Hamiltonian dynamics as being driven by the spatial gradient of the Alfv\'{e}n term in the Hamiltonian being purely radial (second of Equations \ref{eq:Hamilton}). The fluid mechanical interpretation for this process is that it represents phase mixing, previously highlighted in LP17, where the shrinkage of scales (growth of the wavevector) arises from decoherence in the fluctuations associated with a spatial variation in the Alfv\'{e}n speed. Once a ray becomes trapped, it is never observed to re-emerge from the field region. This indicates that this process is in some sense irreversible, and in the presence of dissipation, would act as a global energy sink. Note that there is no violation of the topological phase-space segregation of bound vs.~unbounded Hamiltonian orbits: the ``trapping'' referred to here is purely spatial, while the wavevector remains divergent.

The ray trapping phenomenon has its correspondence in the \textsc{snoopy} simulations in the form of high-amplitude arcs tracing the flux surfaces on which ingoing rays are observed to localise. This is apparent in Figure \ref{fig:vy_tFT_w_rays}, which shows examples of both trapped and reflected rays. The two panels correspond to the steep (top) and shallow (bottom) \textsc{snoopy} runs. For the sake of direct comparison, the rays are given initial wavevectors $\mathbf{k}$ equal to the forcing wavevector for the particular \textsc{snoopy} run. The two types of calculations exhibit features that can be associated with one another: where the rays are reflected, there are interference patterns in the underlying fluid velocity field, and where rays are trapped, there are high-amplitude arcs. Notably, it is clear that the ultimate fate of the ray (reflection or trapping) depends on the location where it impinges upon the field region. In the steep case, the rays impinging relatively far from the $x$-axis are the ones that avoid being trapped, while in the shallow case, rays are mostly reflected except for those impinging very close to the $x$-axis. This suggests that the mutual orientations of the incoming rays, the magnetic flux surfaces and the stratification are important for determining the dynamics.

\subsection{Origin of wave trapping/reflection phenomenon}
We shall now attempt to explain this selective trapping/reflection phenomenon using analytic arguments. To simplify the problem, we will consider linearised Boussinesq dynamics in the ideal MHD (i.e.~dissipation-free) regime. The linearised momentum equation is
\begin{align}
  \frac{\del \mathbf{u}}{\del t} = -\frac{\nabla p'}{\rho} + N^2 \theta' \hat{\mathbf{x}} + \frac{\mathbf{F}_\mathcal{L}'}{\rho} \:, \label{eq:lin_mom}
\end{align}
where $\mathbf{F}_\mathcal{L} = (\nabla \times \mathbf{B}) \times \mathbf{B}$ is the Lorentz force, and primes denote the Eulerian perturbation. Under a constant background entropy gradient, the potential temperature is governed by
\begin{align}
  \frac{\del \theta}{\del t} + \mathbf{u} \cdot \nabla \theta = -u_x \:. \label{eq:pot_temp}
\end{align}
Using $\mathbf{B} = \nabla \times \psi \hat{\mathbf{z}} + \lambda \psi \hat{\mathbf{z}}$, we find that
\begin{align}
  \mathbf{F}_\mathcal{L}' = -\nabla \psi' (\lambda^2 \psi + \nabla^2 \psi) - \nabla \psi (\lambda^2 \psi' + \nabla^2 \psi') \:. \label{eq:Lorentz_force_pertb}
\end{align}
In terms of the fluid displacement $\boldsymbol{\xi}$, satisfying $\mathbf{u} = \del \boldsymbol{\xi} / \del t$, and assuming a time-harmonic dependence $\boldsymbol{\xi} \propto \exp(-\rmi \omega t)$, Equation (\ref{eq:lin_mom}) becomes
\begin{align}
  -\omega^2 \boldsymbol{\xi} = -\frac{\nabla p'}{\rho} - N^2 \xi_x \hat{\mathbf{x}} + \frac{\mathbf{F}_\mathcal{L}'}{\rho} \:. \label{eq:lin_mom_xi}
\end{align}
If we invoke the fluid stream function $\Theta$ satisfying $\boldsymbol{\xi} = \hat{\mathbf{z}} \times \nabla \Theta$ and incorporate the ideal MHD condition
\begin{align}
  \Delta \psi = 0 \implies \psi' = -\xi_R \frac{\rmd \psi}{\rmd R} \:, \label{eq:psi_ideal_MHD}
\end{align}
where $\Delta$ denotes the Lagrangian perturbation, we can eliminate $\boldsymbol{\xi}$ and $\mathbf{F}_\mathcal{L}$ in favour of $\Theta$ and $\psi$. Considering the $z$-component of the curl of Equation (\ref{eq:lin_mom_xi}) allows us to eliminate $p'$ and condense everything down to the single scalar equation
\begin{align}
  -\omega^2 \nabla^2 \Theta &= -N^2 \frac{\del^2 \Theta}{\del y^2} + \frac{1}{\rho R} \frac{\del}{\del \phi} \left[ \frac{\lambda^2}{R} \frac{\del \Theta}{\del \phi} \frac{\rmd \psi}{\rmd R} + \nabla^2 \left( \frac{1}{R} \frac{\del \Theta}{\del \phi} \frac{\rmd \psi}{\rmd R} \right) \right] \frac{\rmd \psi}{\rmd R} \:. \label{eq:lin_mom_scalar}
\end{align}

The next step is to consider local Cartesian limits of Equation (\ref{eq:lin_mom_scalar}) in the vicinity of a reference point $(x_0,y_0)$, located at some cylindrical distance $R_0$ from the origin, with polar coordinate $\phi_0$ (angle from the $x$-axis, as per the usual definition). Let us define a rotated set of coordinates $(l, h)$ aligned with the local $\hat{\mathbf{R}}$ and $\hat{\boldsymbol{\phi}}$ directions (respectively) at the reference point. We shall assume that we are far enough away from the origin that curvature of the coordinates can be neglected on the spatial scales in question (true in the short-wavelength limit), so that
\begin{align}
  \frac{\del}{\del l} \to \frac{\del}{\del R} \:, \quad \frac{\del}{\del h} \to \frac{1}{R} \frac{\del}{\del \phi} \:. \label{eq:deriv_trafo}
\end{align}
The new coordinates are related to $x$ and $y$ through a rotational transformation:
\begin{align}
  l &= (x - x_0) \cos \phi_0 + (y - y_0) \sin \phi_0 \nonumber \\
  h &= -(x - x_0) \sin \phi_0 + (y - y_0) \cos \phi_0 \:. \label{eq:rotated_coords}
\end{align}
Under this change of coordinates, assuming that $(h/R_0, l/R_0) = O(\epsilon)$, where $\epsilon$ is small while $\lambda R_0$ is of order unity and $B_h^2/(\rho N^2) = O(\epsilon^2)$, to lowest order in $\epsilon$ Equation (\ref{eq:lin_mom_scalar}) becomes
\begin{align}
  -\omega^2 \left( \frac{\del^2 \Theta}{\del h^2} + \frac{\del^2 \Theta}{\del l^2} \right) &= -N^2 \cos^2 \phi_0 \frac{\del^2 \Theta}{\del h^2} - N^2 \sin (2\phi_0) \frac{\del^2 \Theta}{\del h \del l} \nonumber \\
  \quad - N^2 \sin^2 \phi_0 \frac{\del^2 \Theta}{\del l^2} &+ \frac{1}{\rho} \left[ \frac{\del^4 \Theta}{\del h^2 \del l^2} + \frac{\del^4 \Theta}{\del h^4} \right] \left( \frac{\rmd \psi}{\rmd R} \right)_{R=R_0}^2 \:. \label{eq:lin_mom_trafo}
\end{align}
Here $B_h = -\del \psi/\del R|_{R=R_0}$ and the ordering conditions place it in the strong field regime. Since background variations occur only in the $R$-direction, in the zero-curvature limit the $h$-component of the wavevector would be constant over the local region considered, and separation of the equations in $h$ is exact. Let us then expand in plane waves in the $h$-direction ($\del/\del h \to ik_h$). The above simplifies to
\begin{align}
  &\left( \omega^2 - N^2 \sin^2 \phi_0 - \frac{k_h^2 B_h^2}{\rho} \right) \frac{\del^2 \Theta}{\del l^2} - \rmi N^2 \sin (2\phi_0) k_h \frac{\del \Theta}{\del l} \nonumber \\
  &\quad - k_h^2 \left( \omega^2 - N^2 \cos^2 \phi_0 - \frac{k_h^2 B_h^2}{\rho} \right) \Theta = 0 \:. \label{eq:lin_mom_trafo_2}
\end{align}

Defining the horizontal Alfv\'{e}n speed $v_h^2 \equiv B_h^2/\rho$, Equation (\ref{eq:lin_mom_trafo_2}) then becomes
\begin{align}
  & \left( \omega^2 - N^2 \sin^2 \phi_0 - k_h^2 v_h^2 \right) \frac{\del^2 \Theta}{\del l^2} - \rmi N^2 \sin(2\phi_0) k_h \frac{\del \Theta}{\del l} \nonumber \\
  & \quad -k_h^2 \left( \omega^2 - N^2 \cos^2 \phi_0 - k_h^2 v_h^2 \right) \Theta = 0 \:. \label{eq:lin_mom_trafo_3}
\end{align}
With the $h$-dependence separated out through the $\propto \exp(ik_h h)$ assumption, Equation (\ref{eq:lin_mom_trafo_3}) is effectively a linear, homogeneous, second-order ODE in the coordinate $l$. In the WKB limit, the coefficients can be regarded as approximately constant, so this can be solved by assuming a $\Theta \propto \exp(\Lambda l)$ expansion and forming the auxilliary equation. The roots of this are given by
\begin{align}
  \Lambda = \frac{\rmi N^2 k_h \sin \phi_0 \cos \phi_0 \pm k_h \sqrt{(\omega^2 - k_h^2 v_h^2)(\omega^2 - N^2 - k_h^2 v_h^2)}}{\omega^2 - N^2 \sin^2 \phi_0 - k_h^2 v_h^2} \:, \label{eq:AE_roots}
\end{align}
the inspection of which yields some useful insights, which we will now elaborate upon.

In general, if the roots of the auxilliary equation are purely imaginary, then solutions will be wavelike (propagating). If they are purely real, then the solutions will be exponentially growing/decaying (evanescent). The first term in the numerator of the expression for $\Lambda$ comes from the buoyancy response and is always imaginary, indicating that the solution will always be at least partially propagating. However, the second term in the numerator, which represents the magnetic response, will be either real or imaginary depending on the strength of the field. Outside the field region $v_h^2 = 0$, and given that $\omega^2 \leq N^2$, $\Lambda$ will be purely imaginary. Without a magnetic field the solution is therefore purely propagating.

Let us consider the second term as $v_h^2$ is increased away from zero. The $\omega^2 - N^2 - k_h^2 v_h^2$ factor will always be negative, but the $\omega^2 - k_h^2 v_h^2$ term, which is initially positive, will eventually reach zero when the field becomes sufficiently strong. This corresponds to encountering the critical surface ($\omega^2 = k_h^2 v_h^2$). Past this point (i.e.~where $k_h^2 v_h^2 > \omega^2$), the solution will acquire an evanescent component. The smaller the value of $\omega$ compared to $N$, the greater the magnitude of the second (evanescent) term compared to the first (propagating) term and the more strongly decaying the solution with respect to space. Physically, it predicts the expulsion of wave action from the field region when the resonance point is crossed. This explains the low velocity amplitudes within the critical surfaces seen in Figures \ref{fig:steep_vy_tFT} and \ref{fig:shallow_vy_tFT}, and also the increasing effectiveness of the observed expulsion with increasing field strength.

The answer to why two distinct types of behaviour (trapping vs.~reflection) occur lies in the denominator of Equation (\ref{eq:AE_roots}). If $\omega^2 < N^2 \sin^2 \phi_0$, then the denominator will always be negative for any value of the field strength. However, if $\omega^2 > N^2 \sin^2 \phi_0$, then as the field strength increases away from zero, at some point, namely when
\begin{align}
  \omega^2 - N^2 \sin^2 \phi_0 = k_h^2 v_h^2 \:, \label{eq:divergence_cond}
\end{align}
the denominator will reach zero and the wavenumber of the solution will diverge. This occurs at a lower field strength than that satisfying the resonance criterion, and so we should expect the divergent regions to be located outside the critical surfaces. Specifically, divergent (large) wavenumbers will be expected in the region where
\begin{align}
  \sin^2 \phi_0 < \sin^2 \phi_* \equiv \frac{\omega^2}{N^2} \:. \label{eq:trapping_angle}
\end{align}
For reasons explained below, we shall refer to this region as the \textit{trapping region}, and $\phi_*$ as the \textit{trapping angle}. It can be seen in Figure \ref{fig:trapping_angles}, which shows the \textsc{snoopy} velocity fields overlaid with the predicted trapping regions (blue lines), that there is a close correspondence between the trapping angle and the angle subtended by the high-amplitude arcs. This is particularly clear in the top row (steep run). The trajectories of trapped rays, overlaid in red, exhibit an even closer correspondence with the predicted trapping regions, but this is somewhat less surprising given that Equation (\ref{eq:trapping_angle}) was derived under the WKB approximation.

We conclude that the trapping phenomenon arises from a singularity in the Boussinesq equations that appears when a magnetic field is present, and applies to a frequency-dependent range of angles between the field lines and the stratification. We shall refer to Equation (\ref{eq:trapping_angle}) as the \textit{trapping condition}. It is a curious result that, as we shall discuss in \S\ref{sec:discuss}, may have important physical consequences relevant for the damping of global modes possessing gravity-mode character. Some physical insight can be gleaned by restoring derivatives with respect to $h$ and recasting the divergence condition Equation (\ref{eq:divergence_cond}) in the form
\begin{align}
  (\omega^2 - N^2 \sin^2 \phi_0) \Theta + \frac{B_h}{\rho R_0} \frac{\del}{\del h} \left( R_0 B_h \frac{\del \Theta}{\del h} \right) = 0 \:, \label{eq:spher_phasemix}
\end{align}
which can be regarded as an equation of motion for fluid elements at the singularity. The resemblance between the last term on the LHS and the torsional operator in equation (15) of LP17 is to be noted. The presence of the additional buoyancy term, compared to equation (14) of LP17, reflects the fact that this describes spheroidal rather than torsional motions. Nonetheless, we see that the dynamics at the singularity and those of the torsional Alfv\'{e}n modes studied in our previous work are closely related. This, and the knowledge that a cascade to arbitrarily small scales occurs at the singularity, points to an association of the trapping regions with phase mixing, the process responsible for the damping provided by the torsional Alfv\'{e}n resonance mechanism.

As a comment, we remark that the critical latitude singularity observed in studies of the rotation problem, where rays reflecting tangentially off an internal surface tend to focus into a high-amplitude beam \citep{Ogilvie2009, Rieutord2010}, is not seen in the \textsc{snoopy} simulations. The reason for this is likely to be that the critical latitude in fact coincides with the trapping angle, and so the waves dissipate within the field region rather than being permitted to form an external beam.

\subsection{Torsional Alfv\'{e}n resonances}
In LP17, we discussed the resonant excitation of torsional modes through coupling with spheroidal modes. The analogous process here is the coupling of motion in the $x,y$-(``meridional'') plane into motions in the $z$-(``azimuthal'') direction. Axisymmetric torsional modes take the form of standing Alfv\'{e}n waves on magnetic flux surfaces, which for the cylindrical Prendergast model can be shown to be simple sinusoids (since the Alfv\'{e}n speed is constant on flux surfaces) with an integer number of wavelengths around the circumference. The associated fundamental vibration frequency $\omega_A(R)$ is given simply by the Alfv\'{e}n speed $v_A(R)$ divided by the circumference, and multiplied by $2\pi$. Each flux surface supports in principle an infinite number of modes corresponding to the spectrum of vibrational harmonics, with frequencies that are integer multiples of the fundamental. If the spheroidal mode can be regarded as an external driver, one expects flux surfaces on which there exist torsional modes whose frequencies match that of the spheroidal mode to resonate with that mode.

\begin{figure*}
  \centering
  \includegraphics[clip=true, trim=1cm 0cm 1cm 0cm, width=\textwidth]{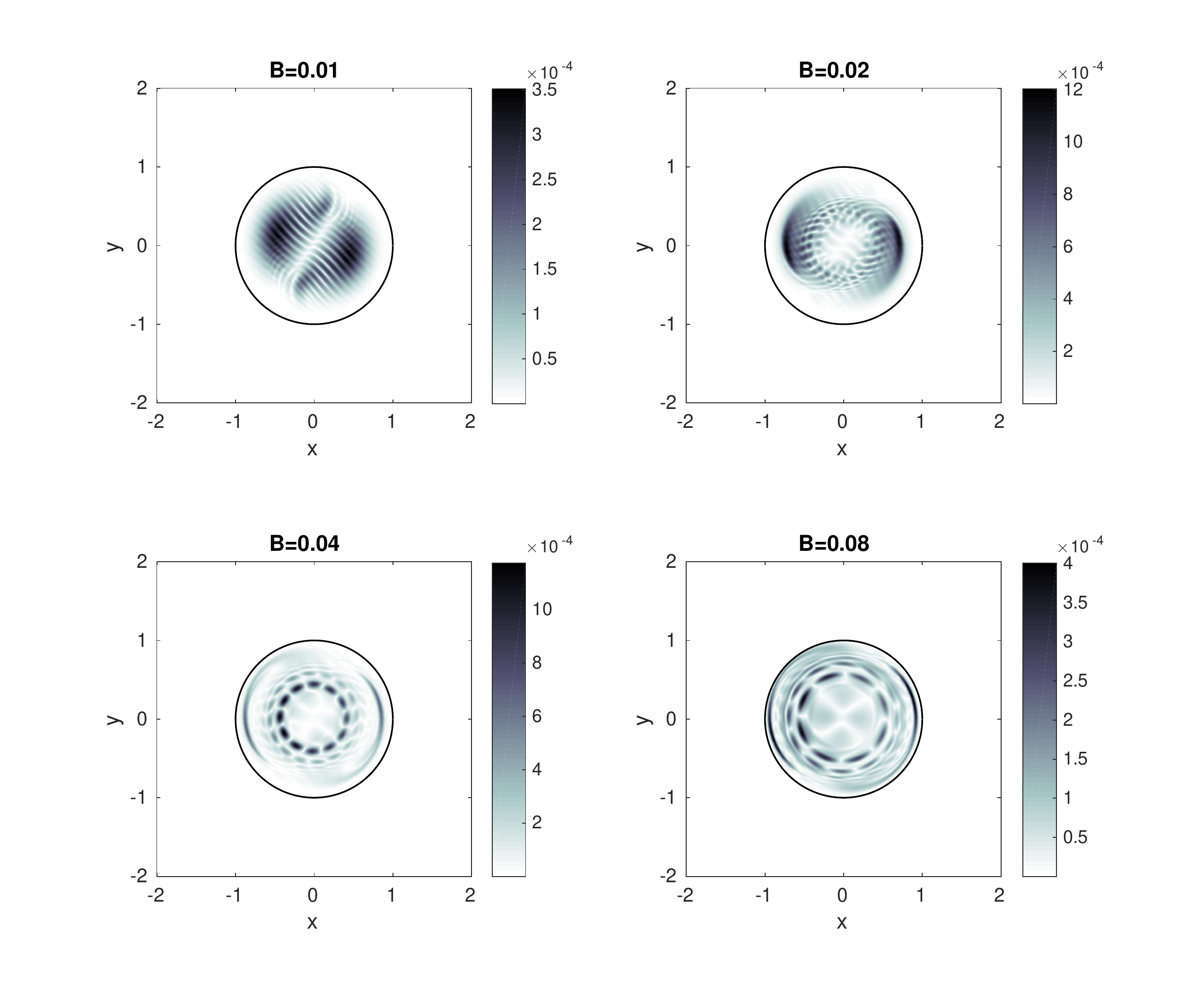}
  \caption{Spatial distribution of the temporal Fourier amplitudes of the $v_z$ component, at four different values of the field strength, for the \textsc{snoopy} run with $[i,j,k]=[15,20,0]$ (steep) at the forcing frequency $\omega_f = 0.8$. The Fourier transform is taken over the interval $200 < t < 1000$. The field region boundary is indicated by the black circle. Critical surfaces are present for all except the top-left panel ($B=0.01$).}
  \label{fig:vz_tFT_i15_j20_k0}
\end{figure*}

The \textsc{snoopy} simulations indeed display evidence for this resonant excitation. Fourier plots of the $z$-component of the velocity, visualised in an identical manner to Figures \ref{fig:steep_vy_tFT} and \ref{fig:shallow_vy_tFT}, are shown in Figure \ref{fig:vz_tFT_i15_j20_k0} for a selection of field strengths. At low field strengths (no critical surfaces), the $v_z$ patterns tend to be slaved to the underlying spheroidal velocity field. This is consistent with the magnetic field having less dynamical importance compared to buoyancy effects: the low Alfv\'{e}n speed compared to the gravity wave phase speed means the magnetic field has no time to produce its own response. However, for strong fields, the $v_z$ patterns organise themselves along magnetic flux surfaces (concentric circles), indicating the acquisition of a substantial Alfv\'{e}nic character. We observe, particularly for the lower two panels, that motions tend to develop on discrete radial surfaces. We interpret these to be the resonant surfaces discussed above.

\begin{figure*}
  \centering
  \includegraphics[clip=true, trim=1cm 0cm 1cm 0cm, width=\textwidth]{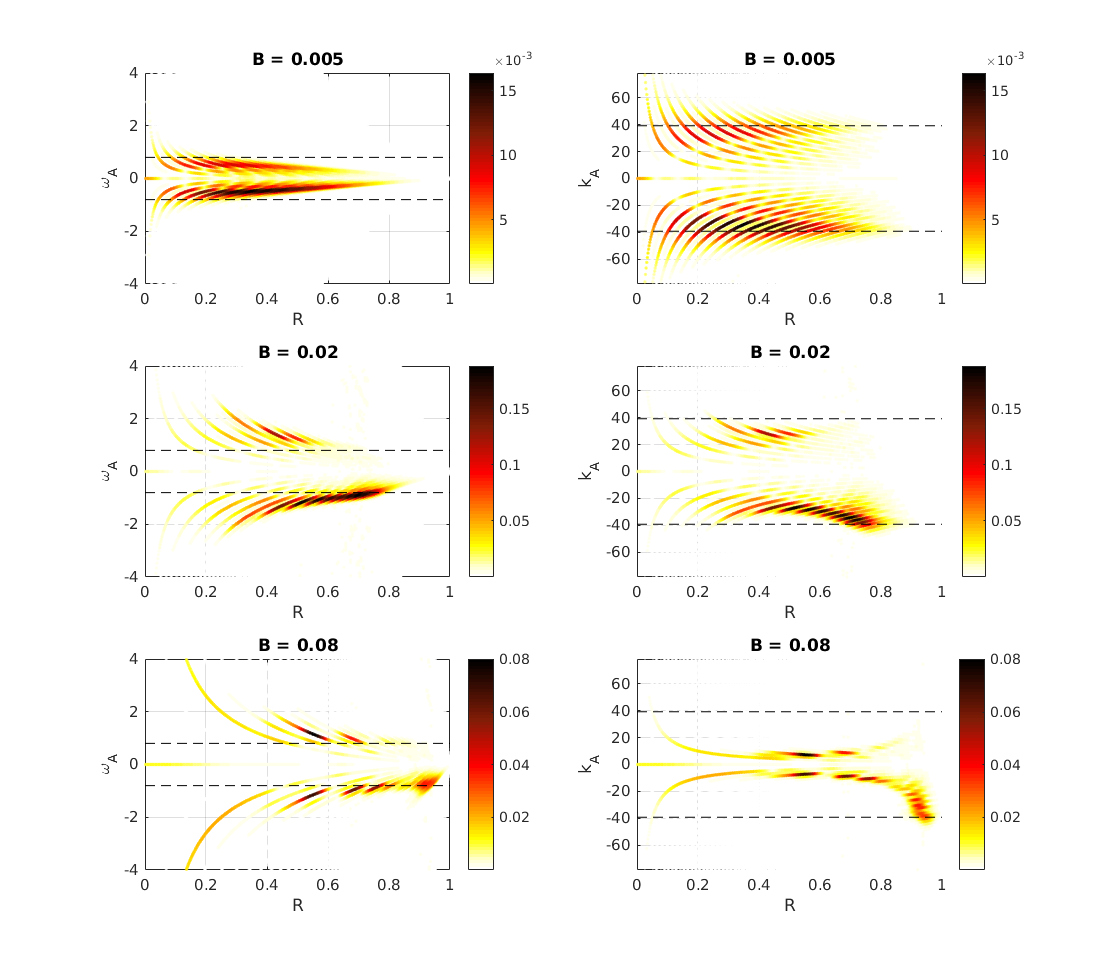}
  \caption{Absolute values of the coefficients of eigenfunction decomposition of the $v_z$ response (e.g.~Fig.~\ref{fig:vz_tFT_i15_j20_k0}) into torsional Alfv\'{e}n modes inside the region of non-zero magnetic field, as a function of radial distance. These are for the \textsc{snoopy} run with $[i,j,k]=[15,20,0]$ (steep), at the value of the forcing frequency, over the interval $200 < t < 1000$. The three rows correspond to three different field strengths, the lower two for which critical surfaces exist. In the left column the vertical axis corresponds to the associated Alfv\'{e}n frequency $\omega_A = k_A v_A$, while in the right column the vertical axis corresponds to $k_A = \mathcal{J}/R$ where $\mathcal{J} \in \mathbb{Z}$ is the harmonic index. Dashed lines indicate the forcing frequency $\omega_f$ (left column) and the forced wavenumber (right column) for comparison.}
  \label{fig:aj_i15_j20_k0}
\end{figure*}

A means of quantifying the properties of the $v_z$ patterns is through a Fourier analysis localised to magnetic flux surfaces. That is, for each value of $R$, we extract the $v_z$ values from the \textsc{snoopy} simulations and then perform a Fourier transform as a function of time and distance along the circumference (azimuthal direction). The resulting Fourier amplitudes are analogous to the $a_j$ coefficients introduced in LP17, which give an indication of the coupling strengths between the driving pattern and each torsional mode. These are shown plotted in Figure \ref{fig:aj_i15_j20_k0} for a selection of field strengths from the steep run. What we find is that for weak fields (top row), the modes that are preferentially excited are those that have wavelengths close to that of the forcing, even if their natural frequencies are significantly discrepant with the forcing frequency. In this regime, the weakness of the magnetic field renders it dynamically unimportant, and so motions are slaved to the driving pattern (Alfv\'{e}n speeds are not sufficient for the Lorentz force to back-react).

In contrast, for strong fields (two bottom rows), it is the modes whose natural frequencies match that of the forcing that are most effectively excited. We interpret this as evidence of resonant interaction with the field. There is an apparent mismatch in spatial scales, but this can be attributed to the strong modification of incoming waves by the magnetic field: on close inspection, the spatial scales of the $v_x$ and $v_y$ patterns found inside the critical surfaces are considerably larger than those outside, and they increase with increasing field strength. Physically, this can be understood as the magnetic field providing an additional restoring force for the waves, boosting their phase speed at a given wavenumber. To maintain the same frequency, the wavelength must lengthen (cf.~the passage of light from a medium of high to low refractive index). Thus in actuality, the wavenumbers measured for the torsional modes more closely match those of the underlying spheroidal motions than na\"{i}vely suggested by Figure \ref{fig:aj_i15_j20_k0}.

In LP17, we identified phase mixing as an important physical process for dissipating energy from torsional Alfv\'{e}n waves excited resonantly by spheroidal gravity modes. The net effect is to damp the spheroidal motions at a rate that depends only on the efficiency of spheroidal-torsional mode coupling. While our results appeared to yield damping rates of a level consistent with observations, an unverified assumption was that the structure of the spheroidal modes was unaffected by the magnetic field at the strengths required for the mechanism to be efficient. The \textsc{snoopy} simulations have provided a straightforward means of checking this, a task that is analytically non-trivial to accomplish. They show that although spheroidal-torsional resonant coupling seems to be occurring, this is unlikely to be effective as a means of damping the spheroidal modes in the context of the dipole dichotomy problem. The reason for this is the expulsion of wave action from the regions within the critical surfaces, leading to very low amplitudes of the driving motions in the strong-field regions where the potentially large number of torsional resonances reside. The physical origin of this expulsion/attenuation is discussed above in the analysis leading to Equation (\ref{eq:AE_roots}). Importantly, this suggests that our calculation of damping rates in LP17, based on the assumption that the structure of the spheroidal modes is unaffected within the field region, is invalid in the regime of field strengths where it might otherwise be effective. The quantitative impact on the associated damping rates, if given by the square of the reduction factor in wave amplitudes inside and outside the critical region, would be to reduce these by around four orders of magnitude. This suggests that the torsional Alfv\'{e}n resonance mechanism may have problems in accounting for the observations.

\section{Discussion}\label{sec:discuss}

\subsection{Relevance to dipole dichotomy problem}
Recent claims of the existence of strong (i.e., dynamically significant) magnetic fields in red giant cores as the explanation for anomalously high dipole mode damping rates \citep{Fuller2015} are at present supported only by indirect evidence, namely that this phenomenon is restricted to stars massive enough to have previously hosted core dynamos when on the main sequence \citep{Stello2016}. Although this link appears sufficiently compelling for the idea to deserve further attention, the physics of the interaction between gravity waves and magnetic fields strong enough to modify their dynamics remains poorly understood.

Our results so far suggest that although the mechanism proposed in LP17 may not be efficient enough to account for the observations of depressed dipole modes in red giants, there is a closely-related process that may achieve this. Rather than a channelling of wave energy from the input motions (spheroidal in nature) into the torsional modes, followed by a loss of energy through phase mixing, our simulations indicate that a loss of energy may occur more directly through phase mixing of the Alfv\'{e}n-modified spheroidal modes themselves. The occurrence of this process relies on the presence of a sufficiently strong magnetic field, but this is not the only criterion. In addition, the orientation of the magnetic field with respect to the stratification is important as it determines the fate of the incoming waves, i.e.~whether they will be reflected or trapped. This property of the system has not previously been appreciated, and could explain why the data indicate only a fractional loss of energy within the g-mode cavity.

In \S\ref{sec:trapping_reflection} we derived a criterion for the trapping angle as being $\sin^{-1} (\omega/N) \approx \omega/N$ in the low-frequency limit (note that $N/(2\pi) \gg \nu_\text{max}$ in typical red giant cores). Though at first glance this appears specific to our cylindrical geometry, our neglect of all terms associated with geometric curvature of the coordinates implies that the result is a local one that generalises to any topologically equivalent system (including spherical stars, as long as one considers the short-wavelength limit). The trapping criterion is to be interpreted as a condition on the local angle between the flux surfaces and the stratification, and defines a set of regions upon which an impinging wave would experience complete absorption. Such regions would subtend some fraction of the surface of the magnetised core. If an oscillation mode has a spatial amplitude function that overlaps with these trapping regions, the mode would experience enhanced damping as a result of this leakage of wave energy. Note that the exact mapping between a given value of $\omega/N$ and the fraction of the core surface occupied by the trapping regions depends on the field configuration in question (we quote values for the spherical Prendergast solution below).

For example, if the trapping regions subtend 5\% of the area of the core, then the associated damping timescale would be 20 times the information crossing time of the star (given roughly by the inverse of the large separation $\Delta \nu$). In general, one would also have to divide through by the square of the transmission coefficient $T$ (which depends on the spherical harmonic degree $\ell$) to account for the fact that only a fraction of wave energy is able to tunnel into the g-mode cavity from where it is input (the convection zone in the p-mode cavity). This is approximately given by the expression
\begin{align}
  T^2 \sim \left( \frac{r_g}{r_p} \right)^{2\sqrt{\ell(\ell+1)}} \:, \label{eq:T2}
\end{align}
where $r_g$ and $r_p$ denote radial coordinates of the upper and lower boundaries of the g-mode and p-mode cavities respectively (lower and upper boundaries of the evanescent region).

A typical red giant might have $\nu_\text{max} \sim 100 \upmu$Hz and $N/(2\pi) \sim 10$ mHz in the core, implying a trapping angle of $\phi_* \sim 0.01$. For the spherical Prendergast field solution considered in LP17, the corresponding trapping regions subtend $>$13\% of the area in the outer 5\% of the core, which is where flux surfaces are closest to being horizontal (i.e.~aligned with the stratification). As can be seen in Figures \ref{fig:steep_vy_tFT} and \ref{fig:shallow_vy_tFT}, for field strengths just a factor of several larger than the critical, trapping regions may be found in close proximity to the edge of the field region, and so it is possible that in general the trapping might occur quite close to the surface of the core. Let us suppose a characteristic trapping area of $\sim$10--20\%. Values of $T^2$ estimated for the red giant model in LP17 for a mode near $\nu_\text{max}$ are 0.2, $8 \times 10^{-3}$ and $2 \times 10^{-4}$ for $\ell$ = 1, 2 and 3, respectively. Typical large separations for a red giant are $\Delta\nu \sim 10 \upmu$Hz, implying an information crossing time of $\sim$1 day. Damping times then evaluate to $\sim$20--50 days for $\ell = 1$, $\sim 10^3$ days for $\ell = 2$, and $\sim 10^4$ days for $\ell = 3$. In the case of the dipole modes, this process is thereby predicted to generate damping rates rivalling those associated with convection. If in operation, it could produce measurable depressions of the dipole mode visibilities: the above values yield $v_1 \sim$ 0.6--0.8, with $v_1$ the dipole visibility normalised in the manner of \citet{Mosser2017a}. For quadrupole, octupole and higher-multipole modes the damping becomes successively smaller owing to the $\ell$-dependence of the transmission coefficient.

The above calculation assumes that the mode samples the entire surface of the magnetised core. It is to be noted that the spatial amplitude functions of the (well-known) regular oscillation modes are relatively localised in space. However, in contrast, chaotic modes are ergodic and thereby have spatial amplitude functions that in principle probe the entire surface of the core. This suggests that chaotic modes of a given frequency should undergo larger damping rates due to this leakage process than their regular counterparts (the restricted nature of the spatial amplitude functions of regular modes gives them a greater chance of ``avoiding'' the trapping regions). As field strengths increase, one expects an increasingly large fraction of g-modes to be chaotic, and so this assumption should be justified in the strong-field limit. For the g-modes that remain regular, we predict that the non-isotropic distribution of the trapping regions (assuming that the field has a simple, large-scale geometry) would produce an orientation-dependent damping rate. That is, different $m$-values of a rotational multiplet might be expected to exhibit different levels of mode depression, depending on the field configuration and its orientation with respect to the rotation axis.

Equation (\ref{eq:trapping_angle}) predicts a dependence of the leakage area on the ratio $\omega/N$. This ratio is expected to decrease as stars evolve along the red giant branch, since continued contraction of the core leads to an increase in $N$ and the expansion of the envelope leads to a decrease in $\omega$. Assuming that the field configuration does not dramatically change over an evolutionary timescale, the trapping regions will shrink as the star evolves. Older red giants are therefore expected to experience smaller rates of damping due to this process, and hence exhibit larger dipole visibilities. This is qualitatively consistent with the observations \citep[e.g.][fig.~3a]{Mosser2017a}, which show an increase in $v_1$ with decreasing $\Delta \nu$ (smaller $\Delta \nu$ values indicate a larger/older star).

\subsection{Relevance to solar interior}
The stably stratified interior of the Sun is separated from the convective envelope by a thin layer (the tachocline, believed to be the seat of the solar dynamo), where strong shear leads to amplification of the background magnetic field. Gravity waves excited by overshooting at the base of the convective envelope must meet this magnetised layer as they propagate downwards into the core. The influence of the tachocline field on gravity waves has been studied by a number of authors \citep{Schatzman1993, Burgess2004, Rogers2010, Rogers2010a, MacGregor2011, Mathis2011, Mathis2012}, in an effort to assess the viability of wave-induced angular momentum transport between the convection zone and solar core. While the above works treat the case of a purely horizontal magnetic field, our results generalise to other orientations. We draw similar conclusions in that for fields of dynamically important strengths, significant hindrance to gravity wave propagation occurs.

A point to be noted is that while we have established that a field of any strength will induce significant interactions with gravity waves in some part of phase space, it is conversely true that for a field of any strength, there will be some part of phase space where propagation will not be adversely affected. The overall consequence for the solar core rotation problem will therefore depend on the spectrum of gravity waves excited at the base of the convective envelope, as well as the magnitude/distribution of tachocline fields, which is not well known \citep[e.g.][]{Barnabe2017}.

In the above context detection of long period g-modes has been reported by \citet{Fossat2017}, who infer that the core is spinning four times faster than the envelope. We comment that the trapping phenomenon and associated phase mixing damping process found here may well provide an additional energy loss route for solar g-modes. For modes below a certain frequency (depending on the tachocline field), this would act to reduce surface amplitudes compared to current theoretical estimates. It would also be expected that the most adversely affected modes would not possess a regularly period-spaced spectrum, but rather these would occupy a chaotic (irregular) regime. This might have implications for data analysis if one attempts to identify solar g-modes based on period regularities.

\subsection{Relevance to compact star oscillations}
Many degenerate stars (such as white dwarfs and neutron stars) are stably stratified throughout most of their volume, permitting the existence of g-mode oscillations \citep[e.g.][]{Reisenegger1992}. Non-radial spheroidal oscillations such as these can generate gravitational radiation, offering a means of probing the interiors of compact objects \citep{Andersson2011, Glampedakis2017}.

A significant fraction of compact stars are known to be magnetised at some level. Our results show that strong interactions with gravity waves can occur regardless of the relative values of gas and magnetic pressures, and so it is likely that at least some of the g-mode oscillations in compact objects will be modified in spatial structure by a magnetic field. This may have implications for the efficiency with which g-modes are excited, for example through nonlinear coupling to r- and f-modes which can in turn grow due to the Chandrasekhar-Friedman-Schutz instability \citep{Friedman1975}. Furthermore, the enhanced damping provided by the trapping phenomenon for modes in affected regions of phase space is likely to shorten the decay times of these modes. The phenomenon of mode chaos is also expected to apply in this regime, implying the existence of an irregular component to the spectrum.

Modifications to the structural and frequency properties of various types of modes in compact stars by a magnetic field, such as r-modes which are likely to generate significant gravitational radiation \citep{Papaloizou1978}, have been investigated extensively via normal mode analyses invoking finite-series spherical harmonic expansions \citep{Morsink2002, Lee2005, Lee2007, Lee2008, Lee2010, Asai2014, Asai2015, Lee2018a, Lee2018}. Such approaches could be complemented by ray calculations of the type described here, offering insight into the local-scale physics of associated wave interactions, and may be a possible course of future work.

\subsection{Limitations}
Without solving the full eigenvalue problem, it is not straightforward to predict the properties of the frequency spectrum in the strong-field regime. The most that can be said at the current stage of investigation is that the spectrum is likely to have a significant irregular component, i.e.~the familiar notion of regular period spacings ceases to apply. The transition to chaotic dynamics implies that spacings between g-modes will instead follow a stochastic distribution with properties that can only be described in a statistical sense. While patterns may still exist in the subset of modes that remain regular, the fraction of such modes decreases with increasing field strength, and so in practice these would be difficult to identify. Previous work on the rapid rotation problem has had some success with applying the Weyl formula \citep{Weyl1912}, which has its origins in quantum mechanics, to estimating the average density of p-modes. Below is a cursory discussion of its extension to the magnetised g-mode problem.

The density of states as a function of frequency can be regarded as the sum of two terms, one describing the average density (given by the Weyl formula) and the other describing fluctuations about it. The leading term of the Weyl formula, which dominates in the high-frequency limit, can be obtained from general principles if one assumes that each mode occupies on average a fixed volume of phase space. This is given by $(2\pi)^\mathcal{N}$, where $\mathcal{N}$ is the dimensionality of the system. Then the number of modes having frequencies less than some given value $\omega$ is equal to the volume of phase space for which $H(\mathbf{x}, \mathbf{k}) < \omega$, divided by $(2\pi)^\mathcal{N}$. This may be understood if one imagines dividing the spatial volume up into infinitesimal cubes, and assuming that in each of these the modes can be taken to be a complete set of plane waves satisfying periodic boundary conditions. In the spherically symmetric g-mode case, the Hamiltonian takes the form given by Equation (\ref{eq:gravity_DR}). Rather than consider the region where $H(\mathbf{x},\mathbf{k}) < \omega$, as is typical in quantum mechanics, for the g-mode problem it is more meaningful to consider instead the region where $\omega < H(\mathbf{x},\mathbf{k}) < N$ (note that an infinite volume of phase space, and so number of g-modes, exists for $H(\mathbf{x},\mathbf{k}) < \omega$).

Restricting ourselves to specified values of $m$ and $\ell$, the associated phase-space volume is given by the integral over the region where
\begin{align}
  |k_r| < \frac{\sqrt{\ell(\ell+1)}}{r} \left( \frac{N^2}{\omega^2} - 1 \right)^{1/2}
\end{align}
and $r_1 < r < r_2$, where $r_1$ and $r_2$ are the turning points of the g-mode cavity, i.e.~radial coordinates where $\omega = N$. This predicts the number of g-modes with frequencies greater than $\omega$ to be
\begin{align}
  n_g(\omega) = \frac{1}{\pi} \int_{r_1}^{r_2} \frac{\sqrt{\ell(\ell+1)}}{r} \left( \frac{N^2}{\omega^2} - 1 \right)^{1/2} \rmd r \:. \label{eq:n_g}
\end{align}
One recognises the right-hand side of the above expression as being $n - 1/2$, where $n$ is the radial order, a result known from WKB analysis of the fluid equations \citep[e.g.][]{Gough2007}. Since for fixed $\ell$ and $m$ the radial order essentially counts the number of g-modes above frequency $\omega$, we see that the (g-mode equivalent of the) leading order term of the Weyl formula gives a result consistent with what is already known from asymptotic theory.

When a magnetic field is imposed, the effect on the Hamiltonian is to add an extra Alfv\'{e}n frequency term. It is unclear as to whether the addition of this term will produce a systematic change in the volume of phase space associated with $\omega < H(\mathbf{x},\mathbf{k}) < N$, compared to the unmagnetised case. For any given frequency interval, the Alfv\'{e}n term will shift some parts of phase space in and others out, with details depending on the field configuration and $N(\mathbf{r})$ profile. Suffice to say that the leading term of the Weyl formula alone does not predict a systematic change in the average density of modes. Note that we have not considered higher-order terms, which will become more significant at low frequencies. Further to this, sinusoidal fluctuations as a function of frequency about the average density are expected to be induced by periodic orbits of the system, as given by the Gutzwiller trace formula from periodic-orbit theory \citep{Gutzwiller1990} in the quantum mechanical case. Implementation and checking the applicability of an analogue of this formula would require a knowledge of the periodic orbits of the system, in conjunction with a solution of the full eigenvalue problem.

To round off this part of the discussion, we are not aware of any straightforward method/heuristic argument for predicting the effect of a magnetic field on the density of g-modes as a function of frequency (short of solving explicitly for the full eigenspectrum). Phase-space volume considerations alone suggest that either an overall increase or decrease may be possible. Analyses of red giant data by \citet{Mosser2017a} suggest that the density of g-modes does not significantly differ between the two groups of stars. If the dichotomy is indeed a result of strong core fields, then this could be an indication that the volumes of phase space affected positively and negatively by the Alfv\'{e}n term tend to cancel out, yielding approximately the same mode densities (within measurement error).

In \S\ref{sec:survival}, we used \textsc{snoopy} simulations to investigate the possible survival/destruction of normal modes in the presence of a strong field by turning off the forcing and observing how the velocity field subsequently evolved. There appeared to be no obvious differences in the rate of decay between the zero-field and strong-field cases. While this seems to reassure us that modes can still exist in the presence of a strong field, one might wonder why at least some additional damping associated with the trapping phenomenon is not seen. We attribute this to the large information crossing times across the box, which are comparable to the timescales of viscous dissipation, of the order $\sim$100 time units. Given that only a small fraction (corresponding to the fractional trapping area) of energy is lost over each box-crossing time due to this process, the decay is likely to be dominated by viscous dissipation. Note that only a limited range of parameters are usable for the purposes of the \textsc{snoopy} simulations (see discussion in \S\ref{sec:snoopy}), making it difficult to decrease the viscosity enough that this damping process might be isolated. Of course, the combination of parameters used here is not representative of that in an actual star, where the timescale of viscous dissipation far exceeds the information crossing time. This remains a limitation of the numerics, but our analytic arguments and ray calulations nonetheless suggest that the damping arising from the trapping phenomenon should be important in real stellar interiors.

Finally, we acknowledge that our geometry is of a simple idealised form that at first glance differs considerably from a real star. However, as discussed in \S\ref{sec:models}, it possesses many properties which are analogous to the meridional plane section of a star containing an embedded magnetic field. We expect that the phenomena seen here should qualitatively generalise to a full spherical star. There may be additional behaviours that do not reveal themselves here, for example arising from a spatially varying buoyancy frequency or global curvature along the $z$-axis. More detailed simulations in a full spherical geometry would be required to investigate these effects.

\section{Summary \& outlook}\label{sec:summary}
The purpose of this work was to improve the physical understanding of gravity wave interactions with a magnetic field, and to relate this to the existence and properties of global modes in a closed, self-gravitating system such as a star. To summarise, we have established that when fields are of sufficient strengths that the frequencies and spatial scales of Alfv\'{e}n and gravity waves match (critical surfaces exist), significant alteration to gravity wave propagation occurs. For any given field strength, this is always possible in some part of phase space, and the value of the plasma $\beta$ is irrelevant. At or near critical surfaces, gravity waves may either be trapped or reflected. The fate of the waves depends on the wave frequency relative to the buoyancy frequency, and the orientation of the flux surfaces with respect to the stratification. For waves that are reflected, ingoing and outgoing wavenumbers tend to be of the same order, allowing for constructive interference of counter-propagating waves and the formation of global magneto-gravity modes. These modes may be subject to additional damping if part of the wave energy enters trapping regions, whereupon acquisition of strong Alfv\'{e}n character and phase mixing occurs. Since magnetic fields cannot be spherically symmetric, magnetic fields strong enough to influence the structure of a g-mode are expected to cause a transition to dynamical chaos owing to the loss of conserved quantities through symmetry breaking, implying that magneto-gravity modes should form an irregular subset.

We have identified at least three astrophysical scenarios where our results have relevance: these include red giant cores, the solar interior, and compact stars. The trapping/phase mixing process has consequences for the damping of g-modes in selected parts of parameter space in all three of the above systems; in addition, for the solar case there are implications for the efficiency of angular momentum transport. The structural modification of g-modes associated with the appearance of critical surfaces impacts the torsional Alfv\'{e}n resonance mechanism of LP17, and in compact stars, for similar reasons, it would also affect g-mode excitation rates due to nonlinear coupling to other modes. The phenomenon of near-specular reflection has consequences for the ability of stably stratified, magnetised systems to support global magneto-gravity modes, contrary to suggestions that strong magnetic fields should destroy this possibility entirely. Finally, the prediction that g-modes heavily altered by a magnetic field should exhibit chaotic behaviour has important practical consequences, as it provides a separate observational test of whether strong fields might exist inside stars.

There are a number of avenues for future work that may be pursued:
\begin{itemize}
  \item The results of this and other works indicate the breakdown of first-order non-degenerate perturbation theory in predicting the frequencies of magnetically-affected g-modes. It would be useful to identify the point beyond which this breaks down; note that this should occur at weaker field strengths than the transition to chaos.
  \item Given that an observational prediction of strong magnetic fields is mode chaos, a next step might be to implement procedures for detecting the presence of irregular components in the oscillation spectra. Borrowing ideas applied to quantum mechanical systems, one approach to this might be to use the fact that statistically, the spacing distribution of chaotic modes is expected to obey a Wigner distribution \citep{Wigner1967}, whereas that of regular modes obeys a Poisson distribution \citep{Berry1977}. Techniques analogous to recurrence spectroscopy from experimental quantum mechanics \citep{Main1994} might also be useful for searching for signatures of periodic orbits.
  \item As pointed out in the Limitations section, our simulation setup neglects spatial curvature along the azimuthal direction. There may be additional phenomena arising from this that have yet to be identified; such an investigation would require the use of much more computationally intensive, fully spherical global simulations.
\end{itemize}

We end with a cautionary word on how the broader implications of this work are to be interpreted. We and other authors have shown that the interactions of gravity waves with a dynamically significant magnetic field exhibit rich and complicated behaviours that have not previously received much detailed attention, and whose nature is still not thoroughly understood. Although studies of restricted/idealised systems can provide useful physical insights, extrapolation of the results to real stellar interiors must be regarded with caution.

\section*{Acknowledgements}
STL is supported by funding from the Cambridge Australia Trust.











\bsp	
\label{lastpage}
\end{document}